\PassOptionsToPackage{dvipsnames,table}{xcolor}
\documentclass{article}
\usepackage[T1]{fontenc}
\usepackage{preprint,times}

\usepackage{amsmath,amsfonts,bm}

\def\figref#1{figure~\ref{#1}}
\def\Figref#1{Figure~\ref{#1}}

\def\Secref#1{Section~\ref{#1}}

\def\eqref#1{equation~\ref{#1}}

\def\1{\bm{1}}

\DeclareMathAlphabet{\mathsfit}{\encodingdefault}{\sfdefault}{m}{sl}
\SetMathAlphabet{\mathsfit}{bold}{\encodingdefault}{\sfdefault}{bx}{n}

\usepackage{amsthm}

\usepackage{subcaption}

\usepackage{hyperref}
\newcommand{\iconlink}[3]{\raisebox{-0.18em}{\includegraphics[height=1.05em]{figures/icons/#1.pdf}}\,\href{#2}{\textbf{#3}}}
\usepackage{url}
\usepackage{graphicx}
\usepackage{booktabs}
\usepackage{wrapfig}
\usepackage{capt-of}
\usepackage{needspace}

\usepackage{xcolor}
\usepackage{placeins}
\usepackage{flafter}
\usepackage{tikz}
\usepackage{pifont}
\usepackage{listings}
\usepackage{titletoc}
\lstnewenvironment{promptbox}{\lstset{basicstyle=\ttfamily\footnotesize,breaklines=true,breakatwhitespace=false,
  columns=fullflexible,keepspaces=true,tabsize=8,frame=single,framerule=0.3pt,rulecolor=\color{black!40},
  xleftmargin=4pt,xrightmargin=4pt,aboveskip=4pt,belowskip=8pt,showstringspaces=false,upquote=true,
  escapechar=@}}{}

\definecolor{uncertain}{rgb}{0,0.5,0}

\def\Tabref#1{Table~\ref{#1}}

\newcommand{\pc}{\mbox{PreciseCoder}}
\newcommand{\pdb}{PDB}

\newcommand*\circled[1]{\tikz[baseline=(char.base)]{\node[shape=circle,draw,inner sep=0.5pt,font=\scriptsize] (char) {#1};}}
\newcommand{\cmark}{\ding{51}}
\newcommand{\xmark}{\ding{55}}

\title{PreciseCoder: Faithful and Generalizable Debugging Beyond Unit-Test Rewards}
\author{
  \textbf{Yunkai Zhan}$^{1\dagger}$ \quad
  \textbf{Miaosen Chai}$^{2}$ \quad
  \textbf{Shangshang Wang}$^{1}$ \quad
  \textbf{Jike Zhong}$^{1}$ \\
  \textbf{Yuqing Yang}$^{1}$ \quad
  \textbf{Deqing Fu}$^{3,1}$ \quad
  \textbf{Serina Chang}$^{4}$ \quad
  \textbf{Wang Bill Zhu}$^{4\dagger}$ \\
  $^{1}$University of Southern California \quad
  $^{2}$University of Chicago \quad
  $^{3}$Google \\
  $^{4}$University of California, Berkeley \\
  $^{\dagger}$Corresponding authors: \texttt{zhanjame@usc.edu}, \texttt{wangzhu@berkeley.edu} \\
  \multicolumn{1}{@{}c@{}}{\makebox[\textwidth][c]{\normalfont\rule{0pt}{1.6em}%
    \iconlink{globe}{https://precise-coder-rl.github.io}{Webpage}\qquad
    \iconlink{github}{https://github.com/precise-coder-rl/PreciseCoder}{Code}\qquad
    \iconlink{hf}{https://huggingface.co/PreciseCoder}{Model}}}
}

\preprintfinalcopy

\begin{document}

\maketitle
\lhead{Preprint}

\begin{abstract}
Reinforcement learning (RL) from unit tests is the common recipe for post-training code models, but a unit test only asks whether the final program passes.
We analyze what this unit-test-only reward teaches a model during debugging and find that, while it raises pass rates, it also teaches the model to \textit{edit more} and to \textit{hallucinate bugs} in code that is already correct.
We introduce \textbf{\pc{}}, a training algorithm that augments the unit-test reward with a signal for \textit{semantically verified, localized edits}, which correlates with smaller edits and higher pass rates.
On 4B, 9B, and 27B models, \pc{} reduces the average number of edit lines from 7--15 to 3 at a pass rate on par with unit-test-only training.
Trained only on single-line bugs, \pc{} can generalize to repository-level program repair, where it resolves more issues with smaller patches.
Its precision also transfers from \textit{weaker to stronger} models: given the reasoning trace of a 9B \pc{} model, frontier models such as GPT-5.6 Sol and Qwen3.8-Max gain 4.6--12.4 points in edit precision with no additional training.
Linguistic analysis of the thinking traces shows that our reward encourages the model to reason about individual lines, name the buggy ones, and form consistent bug hypotheses.

\end{abstract}

\section{Introduction}
\label{sec:intro}

\begin{figure}[t]
\centering
    \begin{subfigure}{\textwidth}
        \includegraphics[width=\linewidth]{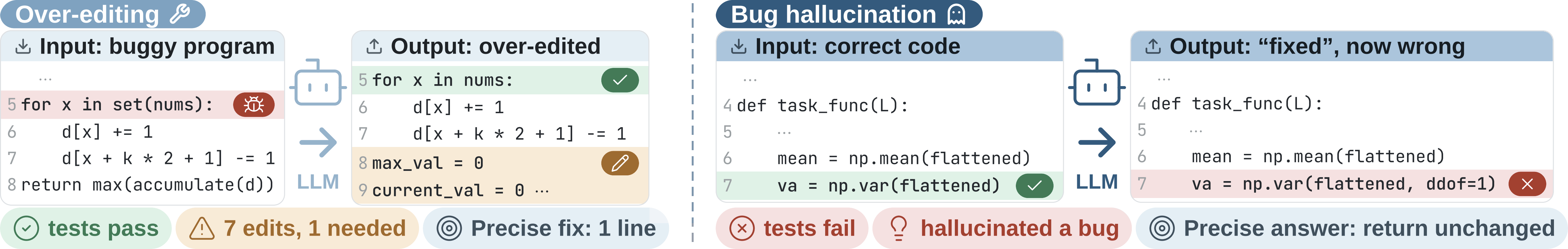}
        \caption{}
        \label{fig:teaser}
    \end{subfigure}
    \begin{subfigure}{\textwidth}
        \includegraphics[width=\linewidth]{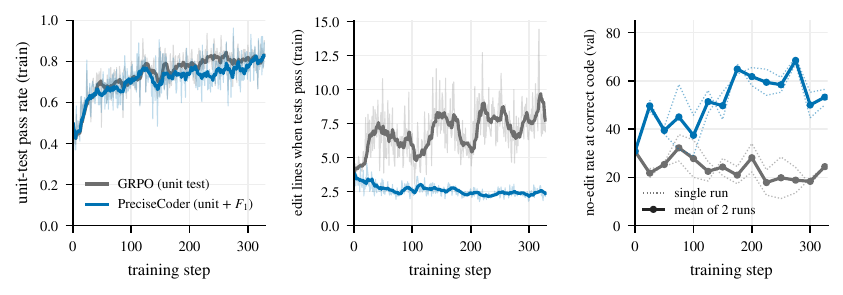}
        \caption{}
        \label{fig:training-dynamics}
    \end{subfigure}
\caption{(a) Illustration of over-editing and bug hallucination phenomenon; (b) Training dynamics of GRPO vs. \pc{} on Qwen3.5-4B, reach the same pass rate (left), but GRPO edits more lines (middle) and leaves fewer bug-free programs untouched (right).}
\label{fig:all_teaser}
\end{figure}

LLMs are increasingly integrated into programs and workflows because they are trained on vast amounts of code. When a developer hands a model a broken function, they want it to identify and repair the fault without modifying code that is already correct. Yet even frontier LLMs rewrite code that was not broken: on \pdb{}~\citep{pdb}, GPT-5.1-Codex passes 76\% of tasks, yet only 40\% of the lines it edits are necessary. This \textit{over-editing} is costly (\Figref{fig:teaser}). Reviewers must inspect every changed line, so the cost of reviewing a repair grows with the size of the edit rather than the size of the bug; even merged pull requests produced by coding agents need human revision 45\% of the time~\citep{claudecodeprs}. Rewrites can also introduce behaviors the tests do not cover, so a patch that passes every test can still regress~\citep{qi2015plausibility,smith2015cure,testoverfitting}.

Reinforcement learning from unit tests has become the dominant approach to post-training code models~\citep{deepseekr1,tulu3,deepswe,kimidev}, so we ask whether it is a root cause of over-editing. 
We first audit what the unit-test reward teaches by tracking the edit statistics of GRPO~\citep{deepseekmath} training rollouts and measuring how each candidate reward term correlates with correctness on the same rollouts. 
We find the unit-test reward raises pass rates without restraining unnecessary edits, and as training proceeds it increasingly makes the model \textit{hallucinate bugs} in code that is already correct (\Figref{fig:training-dynamics}). 
Prior work counters over-editing by penalizing edit distance to the reference~\citep{qimengprepair,editoomuch} or by imitating minimal patches~\citep{paft,adapatcher,recap}, but edit size is only \textit{weakly correlated} with passing the tests, so optimizing it trades pass rate for smaller edits.

We propose \pc{}, which augments the unit-test reward with the semantically verified precision and recall on edits, combined into an $F_1$ score. 
Unlike edit size, this score correlates with both the unit-test outcome and the edit size. 
Hence, the two terms can be optimized together, and we normalize them separately inside the group-relative optimizer. 
Across Qwen3.5-4B, Qwen3.5-9B and Qwen3.6-27B, \pc{} raises \textbf{edit precision} by 24--29\% over the base models and by 26--35\% over unit-test-only GRPO, cuts the \textbf{average edit size} from 7--15 to 3 lines per task, and keeps \textbf{pass rate} on par with unit-test-only GRPO. 
The same reward suppresses bug hallucination although training never shows a correct program: on correct code input, the no-edit (i.e., no operation) rate rises to 47--52\%, against 31--37\% for the base models and 19--31\% after unit-test-only GRPO.

Trained only on single-file single-line bugs, the learned behavior generalizes to complex program repairs. 
On \pdb{}-Wild multi-line and repository-level bugs, precision rises by 20, 17 and 20 points at 4B, 9B and 27B while the edit size shrinks by 63--71\%. 
On SWE-Bench Verified, \pc{} resolves more issues without inflating its patches, and it breaks fewer repositories whose issue is already fixed. 
Trained with agentic RL in a five-turn editing agent, a 9B model reaches 67.4\% edit precision at a 56.3\% pass rate, the best of every 9B model in agentic debugging on both.

Precise debugging also transfers upward, to models we cannot train. Handing the reasoning trace of \pc{}-9B to five stronger receivers, among them GPT-5.6 Sol and Qwen3.8-Max, raises their edit precision by 4.6--12.4 points at their own pass rates, with no weight access, shared tokenizer or training. A linguistic analysis of the traces points to what travels: either reward makes the model think longer and quote the buggy lines, but only \pc{} encourages the model to reason about individual lines, form bug hypotheses, and weigh whether each edit is worth keeping.

\begin{table}[t]
\centering
\setlength{\tabcolsep}{6pt}
\caption{\pc{} vs. representative related work. ``--'' marks evaluation-only studies.}
\label{tab:related}
\resizebox{\linewidth}{!}{
\begin{tabular}{@{}lccccccc@{}}
\toprule
& \multicolumn{2}{c}{Diagnosis} & \multicolumn{4}{c}{Reward signals} & {Generalizability} \\
\cmidrule(lr){2-3}\cmidrule(lr){4-7}\cmidrule(lr){8-8}
& Over- & Bug & Unit & Edit & Location & Partial & Weak-to- \\
& editing & hallucination & tests & extent & verified & fixes & strong \\
\midrule
FixedBench~\citep{fixedbench}     & \xmark & \cmark & --     & --       & --       & --       & \xmark \\
\pdb{}~\citep{pdb}                & \cmark & \xmark & --     & --       & --       & --       & \xmark \\
SWE-RL~\citep{swerl}              & \xmark & \xmark & \xmark & \xmark & \xmark & \xmark & \xmark \\
EA-GRPO~\citep{qimengprepair}     & \cmark & \xmark & \cmark & \cmark   & \xmark   & \xmark  & \xmark \\
OverEdit~\citep{editoomuch}                & \cmark & \xmark & \cmark & \cmark   & \xmark   & \xmark  & \xmark \\
PAFT~\citep{paft}                 & \cmark & \xmark & \xmark & \cmark   & \xmark & \xmark   & \xmark \\
AdaPatcher~\citep{adapatcher}     & \cmark & \xmark & \cmark & \cmark   & \xmark & \xmark   & \xmark \\
RECAP~\citep{recap}               & \cmark & \xmark & \xmark & \cmark   & \xmark   & \xmark   & \xmark \\
\midrule
\pc{}                     & \cmark & \cmark & \cmark & \cmark   & \cmark   & \cmark   & \cmark \\
\bottomrule
\end{tabular}
}
\end{table}

Our contributions are threefold:
\begingroup\setlength{\leftmargini}{1.6em}
\begin{itemize}
\item[\circled{1}] \textbf{Diagnosis.} We show that unit-test-based reinforcement learning raises pass rates while increasing edit size, rewriting correct code, and hallucinating bugs in bug-free programs. We establish this by auditing edit statistics throughout training and measuring each reward term's correlation with correctness and with the number of edited lines when the unit tests pass.
\item[\circled{2}] \textbf{Fix.} We augment the unit-test reward with semantically verified precision and recall. Across 4B, 9B, and 27B models, this reward trains stably, raises edit precision by 24--29 points, reduces the average number of edited lines to approximately three per task without sacrificing pass rate, and curbs bug hallucination, raising the no-edit rate on correct code from 31--37\% to 47--52\%.
\item[\circled{3}] \textbf{Generalization.} The precision carries to multi-line, repository-level and agentic debugging and to SWE-bench Verified, and it travels through the reasoning trace alone: a 9B donor raises the edit precision of frontier models by 4.6--12.4 points. What travels, per a linguistic analysis of the traces, is naming the buggy line and weighing each edit.
\end{itemize}
\endgroup

\section{Related Work}
\label{sec:related}

We review prior work and establish \pc{} as the first approach to incorporate semantically verified, localized edits into RL training for generalizable debugging fidelity (see \Tabref{tab:related}).

\paragraph{Coding models are trained to pass tests.}
Large language models have become widely used coding tools~\citep{humaneval,alphacode}, while agents built on them can resolve real-world repository issues~\citep{swebench,sweagent}.
A common post-training recipe reported by leading AI labs is reinforcement learning with verifiable rewards, using unit-test outcomes as the sole reward signal~\citep{deepseekr1,tulu3,swegym,swesmith,deepswe,kimidev}.
However, the program-repair literature has long established that test-passing patches can overfit~\citep{smith2015cure,tan2016antipatterns}.
Such test overfitting persists in frontier LLMs~\citep{pdb,testoverfitting}, and functional correctness alone fails to capture many qualities developers value~\citep{nofuneval,sweif}.
Moreover, LLM code reviewers often \textit{hallucinate bugs}~\citep{criticgpt}, prompting unnecessary changes to already-correct programs~\citep{selfcorrect,overeager}.
By analyzing training dynamics, we identify unit-test-only rewards as the root cause of both behaviors throughout RL training.

\paragraph{Existing remedies focus on edit size.}
Existing approaches to reducing over-editing share a common design: rewarding smaller edits.
One line of work optimizes edit size alongside functional correctness.
For example, EA-GRPO subtracts a standardized edited-line penalty from the correctness reward~\citep{qimengprepair}, and \citet{editoomuch} subtract excess Levenshtein distance to the reference from an execution reward.
Another line of work separates localization from correction.
RECAP trains a post-hoc refiner using supervised fine-tuning and DPO to shrink generated patches~\citep{recap}, whereas localization-first pipelines constrain where the model can edit~\citep{agentless,locagent,kimidev}.
However, edit-size-based rewards can be difficult to optimize and may \textit{trade test-passing performance} for debugging fidelity.
We instead use semantically verified edit size, enabling smoother convergence and better generalization.

\section{Method}
\label{sec:method}

We describe a debugging task as $\tau=(C,x,T,\Delta^\star)$, where $C$ is a buggy program, $x$ is a specification, $T(\cdot)\in\{0,1\}$ is a test suite with $T(C)=0$, and $\Delta^\star$ is the reference repair that yields the fault-free program.
A policy $\pi_\theta(\hat C\mid C,x)$ emits a revised program $\hat C$; we denote the predicted edit by $\hat\Delta=\mathrm{diff}(C,\hat C)$ and the number of edited lines by $\ell=|\hat\Delta|$.
A standard RL pipeline takes a set of $N$ tasks $\tau_{1..N}$ and optimizes $\pi_\theta$ with the unit-test-only reward $r=T(\hat C)$.
Existing localized debugging approaches use $r=f(T(\hat C),\ell)$, combining test success and edit size linearly or through gating.

\subsection{Fidelity Concerns with Unit-Test-Only GRPO}
\label{sec:symptoms}

We first train Qwen3.5-4B with group-relative policy optimization (GRPO,~\citealp{deepseekmath}) using the reward $r=T(\hat C)$ on PDB~\citep{pdb} tasks, and analyze the training dynamics.

\paragraph{Unit-test-only GRPO learns to pass by editing more.}
As shown by the gray line in \Figref{fig:training-dynamics} (left and middle), the model initially learns to repair programs by editing more: the mean number of edited lines tracks the pass rate, with both rising over steps 0--50 and 100--150 and falling over steps 50--100.
Once the pass rate begins to plateau, however, the reward continues to periodically encourage over-editing, and the mean edit size grows from 3.94 to 8.28 lines on the training set.

\paragraph{The same reward teaches the model to hallucinate bugs.}
Training never shows the model a correct program, yet the effect of the reward on correct programs is measurable.
On a validation set of 200 bug-free programs, the unit-test-only reward teaches the model to hallucinate bugs and modify correct code: the no-op rate on ground-truth programs drops from 30.8\% at the start of training to 20.5\% averaged over the last three evaluations (\Figref{fig:training-dynamics}, right).

\begin{figure}[ht]
\centering
\includegraphics[width=0.55\linewidth]{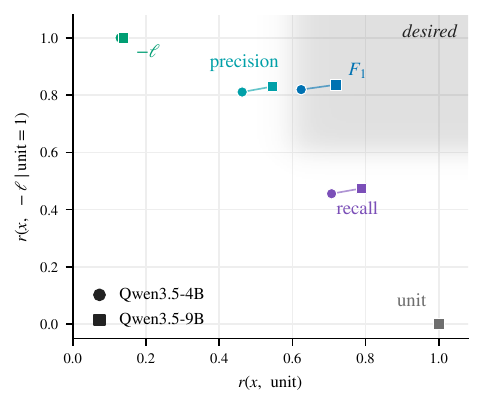}
\caption{Pearson $r$ averaged over all training tasks (circle 4B, square 9B; $n{=}16$ samples per task). $\mathrm{unit}$ is constant when passing and is plotted at $y{=}0$.}
\label{fig:roles}
\end{figure}

\subsection{How Different Reward Terms Correlate}
\label{sec:method:roles}

Next, we analyze how candidate reward components correlate with one another on the Qwen3.5-4B and 9B base models (\Figref{fig:roles}).
Existing methods such as EA-GRPO use linear or gated combinations of the pass indicator $\mathrm{unit}=T(\hat C)$ and the edit size $\ell$.
However, the Pearson correlation between $-\ell$ and $\mathrm{unit}$ is low, at $0.13$ on 4B and $0.14$ on 9B: edit size carries little information about correctness, so optimizing it directly can destabilize training and degrade the pass rate.

To address this, we introduce semantically verified, localized edit rewards.
Inspired by~\cite{pdb}, we score the edit fault by fault: each predicted block $\hat\delta_i\subseteq\hat\Delta$ is aligned to a reference block $\delta^\star_i\subseteq\Delta^\star$, and the model is credited with fault $i$ ($\mathrm{fix}_i=1$) iff the program under the counterfactual edit $\Delta=\Delta^\star-\{\delta_i^\star\}\cup\{\hat\delta_i\}$ passes the tests.
Precision and recall are then
\begin{equation}
  \mathrm{rec}=\tfrac{1}{m}\textstyle\sum_i\mathrm{fix}_i\in[0,1],\qquad
  \mathrm{prec}=\tfrac{1}{\ell}\textstyle\sum_i\mathrm{fix}_i\,
  \min\!\big(|\mathrm{ess}_i|,\,|\delta^\star_i|+\epsilon\big)\in[0,1],
  \label{eq:prf}
\end{equation}
where $m$ is the number of reference faults, $\mathrm{ess}_i\subseteq\hat\delta_i$ is the smallest subset of the predicted block that still repairs fault $i$, and $\epsilon$ tolerates a few extra lines per fault.
Recall asks which faults were repaired; precision asks what fraction of the edited lines were needed to repair them.


We evaluate each candidate reward term by two metrics: its correlation with $\mathrm{unit}$, for \textit{stability}, and its correlation with $-\ell$ among passing samples, for \textit{fidelity} (\Figref{fig:roles}).
$-\ell$ fails the first, and $\mathrm{unit}$, constant among passing samples, fails the second.
$\mathrm{prec}$ correlates well with $-\ell$ (>$0.81$), while $\mathrm{rec}$ correlates well with $\mathrm{unit}$ (>$0.71$).
We therefore design a reward of the form $r=f(\mathrm{unit},\mathrm{prec},\mathrm{rec})$.

First, to avoid spurious block misalignment, we gate $\mathrm{rec}$ on $\mathrm{unit}$: we define $\widetilde{\mathrm{rec}}=1$ if $\mathrm{unit}=1$ and $\widetilde{\mathrm{rec}}=\mathrm{rec}$ otherwise, since a passing program has by definition resolved every injected fault.
We then take $F_1$ as the harmonic mean of $\widetilde{\mathrm{rec}}$ and $\mathrm{prec}$, which reduces to $2\,\mathrm{prec}/(1+\mathrm{prec})$ on passing samples.
$F_1$ correlates well with both $\mathrm{unit}$ (>$0.62$) and $-\ell$ (>$0.82$), the only term in the desired region.

\subsection{Algorithm}
\label{sec:method:optimizer}

We propose \pc{}, which trains with the reward $r_\lambda=(1-\lambda)\,\mathrm{unit}+\lambda\,F_1$.
Note that GRPO standardizes rewards within each group.
Applied to the sum $r_\lambda$, this standardization collapses to the standardized $F_1$ in any group where $\mathrm{unit}$ is constant, so the weight $\lambda$ acts unevenly across groups~\citep{gdpo,mogrpo}.
To mitigate this, we use GDPO~\citep{gdpo}, which standardizes each term within the group before mixing, followed by batch-level rescaling.
Under this normalization, a passing correction retains a larger advantage than a failing one whenever $\lambda<2/(2+\sqrt{2G})$, i.e., $\lambda<0.37$ at $G=6$ (Appendix~\ref{app:ordering}); we use $\lambda=1/3$.
Within this range, the exact value of $\lambda$ matters little: where $\mathrm{unit}$ is constant the advantage equals the standardized $F_1$ regardless of $\lambda$, and elsewhere $F_1$ only breaks ties.

\begin{table}[t]
\setlength{\tabcolsep}{8pt}
\caption{Prior methods trade pass rate for smaller edits; \pc{} keeps pass rate on par with GRPO, while editing $4\times$ fewer lines on \pdb{} Test and, without seeing correct code, hallucinates fewer bugs on GT Debug. Bold: best; underline: second.}
\label{tab:main}
\centering
\resizebox{\linewidth}{!}
{
\setlength{\aboverulesep}{0pt}\setlength{\belowrulesep}{0pt}
\begin{tabular}{@{}lllllll@{}}
\toprule
Methods & \multicolumn{4}{c}{PDB Test} & \multicolumn{2}{c}{GT Debug} \\
\cmidrule(lr){2-5} \cmidrule(lr){6-7}
 & Pass (\%)\,$\uparrow$ & Precision (\%)\,$\uparrow$ & Recall (\%)\,$\uparrow$ & Lines $\downarrow$ & No-op (\%)\,$\uparrow$ & Damage (\%)\,$\downarrow$ \\
\midrule
\rowcolor{blue!12} \multicolumn{7}{l}{\textit{Model: Qwen3.5-4B}} \\
Base         & 36.5 & 35.8 & 48.5 & 15.3 & 30.6 & 19.3 \\
SFT             & 41.2 & 39.7 & 54.7 & 10.2 & 34.9 & 19.4 \\
DPO             & 51.5 & 40.5 & 61.2 & 8.7 & 30.6 & \underline{17.7} \\
GRPO            & \underline{55.1} & 30.0 & 60.1 & 13.8 & 19.3 & 17.8 \\
RECAP             & 38.1 & \underline{59.8} & 59.8 & \textbf{2.9} & 39.6 & 18.2 \\
EA-GRPO         & 53.3 & 58.6 & \underline{70.1} & 3.7 & \underline{43.3} & 18.8 \\
\textbf{PreciseCoder}  & \textbf{55.3\scriptsize{$\pm$0.9}} & \textbf{64.9\scriptsize{$\pm$1.3}} & \textbf{72.9\scriptsize{$\pm$1.1}} & \underline{3.0\scriptsize{$\pm$0.1}} & \textbf{48.0\scriptsize{$\pm$5.8}} & \textbf{13.5\scriptsize{$\pm$1.5}} \\
\midrule
\rowcolor{blue!12} \multicolumn{7}{l}{\textit{Model: Qwen3.5-9B}} \\
Base         & 41.2 & 43.9 & 53.1 & 13.7 & 36.6 & 18.7 \\
SFT             & 47.9 & 50.6 & 63.3 & 7.2 & 41.0 & 15.1 \\
DPO             & 56.6 & 52.1 & 68.6 & 7.3 & 41.3 & 13.4 \\
GRPO            & \textbf{63.9} & 41.7 & 68.4 & 11.9 & 30.7 & \textbf{11.4} \\
RECAP           & 47.6 & 66.1 & 67.0 & \textbf{2.7} & 44.1 & 15.7 \\
EA-GRPO         & 59.3 & \underline{67.0} & \underline{76.0} & 3.5 & \underline{47.0} & 13.5 \\
\textbf{PreciseCoder}  & \underline{62.5\scriptsize{$\pm$1.9}} & \textbf{67.7\scriptsize{$\pm$1.5}} & \textbf{76.8\scriptsize{$\pm$1.1}} & \underline{3.0\scriptsize{$\pm$0.1}} & \textbf{51.6\scriptsize{$\pm$2.7}} & \underline{11.8\scriptsize{$\pm$0.9}} \\
\midrule
\rowcolor{blue!12} \multicolumn{7}{l}{\textit{Model: Qwen3.6-27B}} \\
Base         & 72.9 & 49.8 & 74.1 & 7.4 & 31.1 & 15.5 \\
GRPO            & 76.3 & 43.2 & 71.0 & 7.9 & 26.9 & \underline{12.6} \\
EA-GRPO         & \underline{77.3} & \underline{77.0} & \underline{85.2} & \underline{2.7} & \underline{44.4} & 16.6 \\
\textbf{PreciseCoder} & \textbf{79.3} & \textbf{79.2} & \textbf{86.7} & \textbf{2.6} & \textbf{45.8} & \textbf{11.1} \\
\midrule
\rowcolor{blue!12} \multicolumn{7}{l}{\textit{Frontier Models}} \\
GPT-5.6 Sol (high)      & \textbf{86.1} & \textbf{61.5} & \textbf{87.9} & \textbf{4.1} & \textbf{48.0} & \textbf{9.3} \\
DeepSeek-V4-Pro         & \underline{82.0} & 53.5 & \underline{82.5} & 5.2 & 31.0 & \underline{13.4} \\
DeepSeek-V4.1-Flash     & 76.1 & \underline{57.8} & 80.6 & \underline{4.8} & \underline{39.7} & 21.5 \\
Qwen3.8-Flash-Next      & 75.3 & 55.3 & 78.2 & 5.2 & 22.1 & 17.9 \\
\bottomrule
\end{tabular}
}
\end{table}

\section{Experiments}
\label{sec:experiments}

\subsection{Setup}
\label{sec:exp-setup}

We train on 5{,}274 tasks from \pdb{}-Single-Full~\citep{pdb}, the unfiltered pool of 7{,}589 synthesized single-line-bug programs built on BigCodeBench~\citep{bigcodebench} and LiveCodeBench~\citep{livecodebench} problems.
We test on 1{,}205 held-out tasks from \pdb{}-Single, the verified subset of this pool. The test problems share no base problem with any training task.
We evaluate on unit test passing rate, precision, recall, and edited lines.

In addition, we also test bug hallucination by asking LLMs to fix 1{,}512  ground truth programs, drawn from BigCodeBench and LiveCodeBench problems that do not appear in the above splits.
We evaluate on no operation rate (No-op\%) and damage rate (i.e., test failure rate) from the output.

We use Qwen3.5-4B, Qwen3.5-9B~\citep{qwen2026qwen35} and Qwen3.6-27B~\citep{qwen2026qwen36} as the backbone models, and compare with five baselines: (1) \textbf{SFT}: supervised fine-tuning with rejection-sampling~\citep{editoomuch}, (2) \textbf{DPO}~\citep{rafailov2023dpo}, (3): \textbf{GRPO}: unit-test-only GRPO, (4) \textbf{RECAP}: training a post-hoc edit size refiner after correction~\citep{recap}, and (5) \textbf{EA-GRPO}: penalizing edit size in mostly-passing groups~\citep{qimengprepair}.
We further compare with four frontier models: GPT-5.6 Sol~\citep{openai2026gpt56}, DeepSeek-V4-Pro~\citep{deepseekai2026deepseekv4}, DeepSeek-V4.1-Flash~\citep{deepseekai2026deepseekv41flash} and Qwen3.8-Flash-Next~\citep{qiu2026qwen38next}.
Details in Appendix~\ref{app:implementation} and prompts in Appendix~\ref{app:prompts}.

\subsection{Main Results}
\label{sec:results}

\paragraph{Optimizing edit size trades off pass rate.}
Compared with GRPO in \Tabref{tab:main}, RECAP and EA-GRPO substantially improve fidelity on Qwen3.5-4B: mean edited lines drop from 13.8 to at most 3.7, and precision rises from 30\% to over 58\%.
However, this comes at the cost of pass rate, which falls from 55.1\% to 38.1\% for RECAP and 53.3\% for EA-GRPO.
The same pattern holds on Qwen3.5-9B.
This is consistent with our earlier analysis: because edit size is only weakly correlated with pass rate, optimizing it directly incurs a performance trade-off.

\paragraph{\pc{} improves fidelity while keeping pass rate on par.}
\Tabref{tab:main} reports the mean and standard deviation of \pc{} over 3 runs.
Unlike prior methods, \pc{} reduces mean edited lines from 13.8 to 3.0 on Qwen3.5-4B while keeping pass rate on par with GRPO; the same holds on Qwen3.5-9B.
\pc{} ranks in the top two on every metric, demonstrating its robustness and validating a reward design that balances stability and effectiveness.

\paragraph{\pc{} on Qwen3.6-27B is competitive with frontier models.}
\pc{} on Qwen3.6-27B outperforms its GRPO counterpart, DeepSeek-V4.1-Flash and Qwen3.8-Flash-Next on all measures.
GPT-5.6 Sol attains the highest pass rate (86.1\%) but still over-edits: its precision is 17.7\% below \pc{} (79.2\%), and it edits 1.5 more lines on average than \pc{} (2.6).
This suggests \pc{} could surpass frontier models when scaled to comparable size.

\begin{figure}[t]
\begin{minipage}[t]{0.45\linewidth}
\vspace{0pt}
\centering
\includegraphics[width=\linewidth]{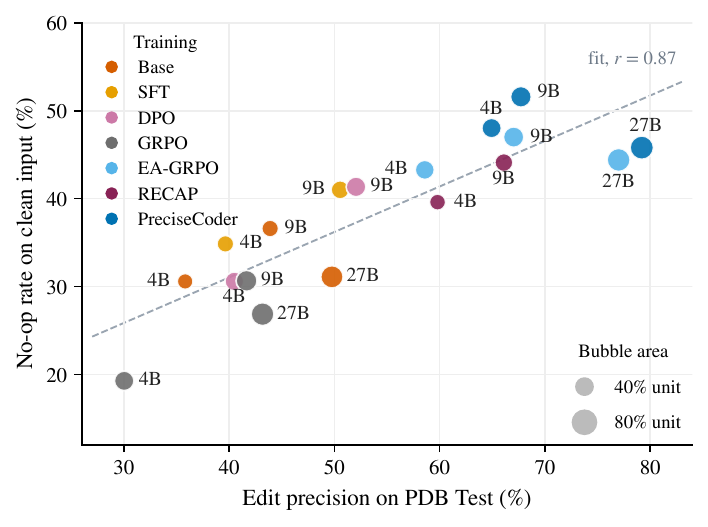}
\caption{No-op rate on GT Debug scales linearly with precision.}
\label{fig:main-results}
\end{minipage}\hfill
\begin{minipage}[t]{0.53\linewidth}
\vspace{0pt}
\centering
\captionof{table}{Reward ablation on Qwen3.5-4B, PDB-Single. $\pm$: std over three runs; $\dagger$: ours; $\ast$: worse than ours by more than the std.}
\label{tab:reward-ablation}
\resizebox{\linewidth}{!}{
\setlength{\tabcolsep}{3pt}
\begin{tabular}{@{}llll@{}}
\toprule
Reward $r$ & Pass & Prec. & Lines \\
          & (\%)\,$\uparrow$ & (\%)\,$\uparrow$ & $\downarrow$ \\
\midrule
$\mathrm{unit}$                                            & 55.1 & 30.0$^*$ & 13.8$^*$ \\
$\mathrm{unit}\cdot\mathrm{prec}$                          & 52.6$^*$ & 63.1$^*$ & 3.3$^*$ \\
$F_1(\mathrm{prec},\mathrm{rec})$                          & 50.0$^*$ & 70.1 & 2.5 \\
$\tfrac23\,\mathrm{unit}+\tfrac13\,\mathrm{prec}$          & 52.3$^*$ & 69.0 & 2.6 \\
$\tfrac23\,\mathrm{unit}+\tfrac13\,\mathrm{rec}$           & 57.2 & 52.2$^*$ & 3.8$^*$ \\
$\tfrac23\,\mathrm{unit}+\tfrac13\,F_1(\mathrm{prec},\mathrm{rec})$ & 52.2{\scriptsize$\pm$1.8}$^*$ & 66.1{\scriptsize$\pm$3.3} & 2.8{\scriptsize$\pm$0.2} \\
\midrule
$\tfrac23\,\mathrm{unit}+\tfrac13\,F_1(\mathrm{prec},\widetilde{\mathrm{rec}})^\dagger$ & 55.3{\scriptsize$\pm$0.9} & 64.9{\scriptsize$\pm$1.3} & 3.0{\scriptsize$\pm$0.1} \\
\bottomrule
\end{tabular}
}

\end{minipage}
\vspace{-4pt}
\end{figure}

\subsection{Bug Hallucination on Correct Code}
\label{sec:gt-debug}

\paragraph{Our reward raises the no-op rate and lowers damage without ever seeing correct code.}
No training row contains correct code and no reward term rewards a no-op, yet restraint against bug hallucination comes for free.
In \Tabref{tab:main}, unit-test-only GRPO lowers the no-op rate from 30.6\% to 19.3\% at 4B, from 36.6\% to 30.7\% at 9B and from 31.1\% to 26.9\% at 27B, whereas \pc{} raises it to 48.0\%, 51.6\% and 45.8\% and breaks fewer correct programs than the base model (damage 13.5\%, 11.8\% and 11.1\% vs.\ 19.3\%, 18.7\% and 15.5\%).
RECAP's refiner and EA-GRPO's size penalty show similar but smaller gains.
Moreover, we show in Appendix~\ref{app:prompt_gt_debug} that \pc{} keeps its advantage on the GT debugging tasks when prompting for restraint~\citep{fixedbench}.

\paragraph{Precision on buggy code and restraint on correct code rise together.}
Plotting the no-op rates in \Tabref{tab:main} against edit precision, we find a near-linear relationship (Figure~\ref{fig:main-results}).
The two skills are highly correlated, with $r=0.87$ across all 4B--27B models ($r=0.94$ within the 4B rows and $0.92$ within the 9B rows).
Conversely, training on a mixture that includes ground-truth debugging tasks also improves debugging fidelity (Appendix~\ref{app:gt_train}).
Either skill thus brings the other along for free.

\begin{figure}
    \centering
    \includegraphics[width=1\linewidth]{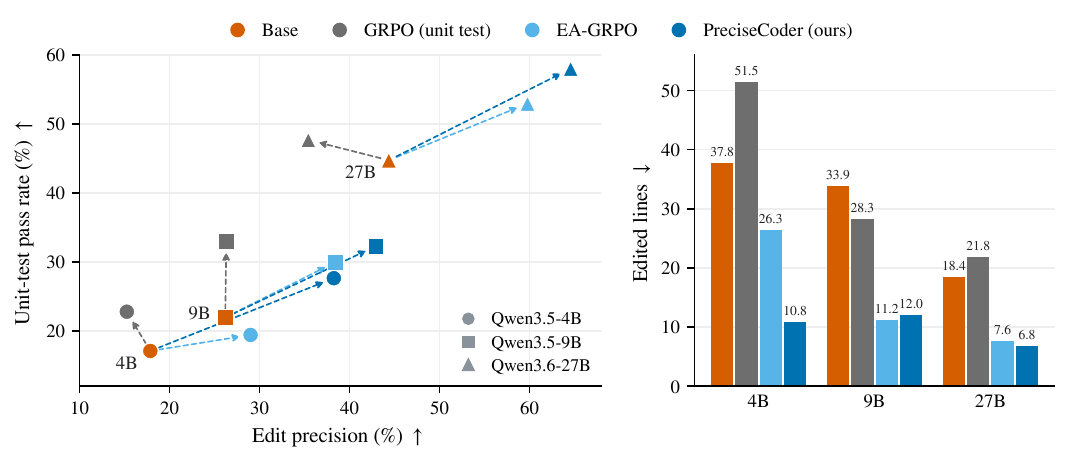}
    \caption{On \pdb{}-Wild multi-line and repository-level repairs, \pc{} matches or outperforms unit-test-only GRPO and EA-GRPO across all metrics.}
\label{fig:pdb-wild}
\end{figure}

\subsection{Reward Ablations}
\label{sec:ablations}

We ablate alternative choices of $f(\cdot)$ in $r=f(\mathrm{unit},\mathrm{prec},\mathrm{rec})$ and find that no single term recovers both fidelity and pass rate: precision alone over-restrains, recall alone under-restrains, and only their gated combination keeps pass rate on par.

\Tabref{tab:reward-ablation} trains Qwen3.5-4B on each reward term alone and in combination.
With the unit test alone, the policy passes 55.1\% of tasks but edits 13.8 lines per task at 30.0\% precision.
Rewarding precision, either as the added term or as the whole reward, yields the most precise repairs in the table (69.0--70.1\% precision, 2.5--2.6 lines) but \textit{passes 3--5\% fewer tasks than ours}.
Rewarding recall alone does the opposite, reaching the highest pass rate (57.2\%) and recall (74.1\%) with the \textit{least precise repairs (52.2\%, 3.8 lines)}, since recall does not charge for lines that repair nothing.
Gating precision on the unit test ($r=\mathrm{unit}\cdot\mathrm{prec}$) does not recover pass rate either: like pure $F_1$, it passes fewer tasks than ours.

\begin{figure}[t]
\begin{minipage}[t]{0.6258\linewidth}
\vspace{0pt}
\centering
\includegraphics[width=\linewidth]{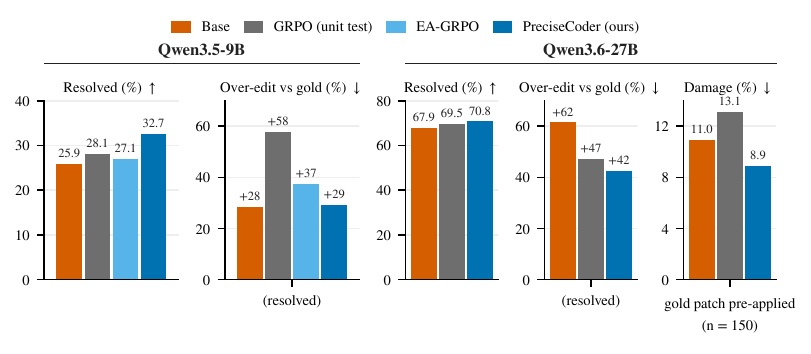}
\caption{\pc{} resolves more issues with smaller patches on SWE-Bench Verified.}
\label{fig:swebench-gt}
\end{minipage}\hfill
\begin{minipage}[t]{0.3492\linewidth}
\vspace{0pt}
\centering
\includegraphics[width=\linewidth]{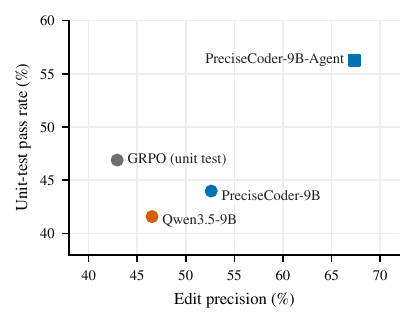}
\caption{Agentic \pc{} closes the multi-turn pass-rate gap.}
\label{fig:agentic}
\end{minipage}
\end{figure}
\begin{table}
\caption{Thinking trace carries correctness and precision for weak-to-strong transfer on \pdb{} Test. \quad $\Delta$: change vs. the receiver's own reasoning.}
\label{tab:transfer}
\centering
\small
\setlength{\tabcolsep}{10pt}
\begin{tabular}{lcccrr}
\toprule
& \multicolumn{3}{c}{With 9B trace} & \multicolumn{2}{c}{$\Delta$} \\
\cmidrule(lr){2-4} \cmidrule(lr){5-6}
Receiver & Pass & Prec. & Lines & Pass & Prec. \\
         & (\%) & (\%) & & (\%) & (\%) \\
\midrule
\multicolumn{6}{l}{\textit{Donor, own trace}} \\
\pc{}-9B            & 64.6 & 67.9 & 3.0 & --- & --- \\
\midrule
\multicolumn{6}{l}{\textit{Own base, prefilled}} \\
Qwen3.5-9B          & 58.3 & 58.4 & 6.5 & \textbf{\boldmath$+$17.1} & \textbf{\boldmath$+$14.5} \\
\midrule
\multicolumn{6}{l}{\textit{Control: untrained Qwen3.5-9B trace}} \\
Qwen3.6-27B         & 66.4 & 53.6 & 7.6 & $-$7.4 & \textbf{\boldmath$+$2.6} \\
\midrule
\multicolumn{6}{l}{\textit{Prefilled, continues thinking}} \\
Qwen3.6-27B         & 78.1 & 62.1 & 5.7 & \textbf{\boldmath$+$4.3} & \textbf{\boldmath$+$11.1} \\
Qwen3.8-Flash-Next  & 75.9 & 60.9 & 6.0 & \textbf{\boldmath$+$0.6} & \textbf{\boldmath$+$5.6} \\
DeepSeek-V4.1-Flash & 74.6 & 63.3 & 4.3 & $-$1.5 & \textbf{\boldmath$+$5.5} \\
\midrule
\multicolumn{6}{l}{\textit{Trace in the user turn}} \\
Qwen3.8-Max         & 82.3 & 68.6 & 3.7 & $-$0.7 & \textbf{\boldmath$+$12.4} \\
GPT-5.6 Sol (high)  & 87.3 & 66.1 & 3.6 & \textbf{\boldmath$+$1.2} & \textbf{\boldmath$+$4.6} \\
\bottomrule
\end{tabular}
\end{table}

\section{Generalization}
\label{sec:generalize}

Training shows \pc{} only single-line faults in single-file programs, repaired in one shot.
We test whether the learned precision survives settings that training never shows, whether it reaches stronger models through the reasoning trace alone, and what the reward changed in that trace.

\subsection{Generalization to Complex Program Repairs}
\label{sec:generalization}

\paragraph{\pc{} generalizes to multi-line and repository-level faults.}
On \pdb{}-Wild~\citep{pdb}, 484 multi-line and repository-level tasks drawn from BigCodeBench, LiveCodeBench and SWE-smith~\citep{swesmith}, \pc{} raises edit precision over the base model by 20.3, 16.7 and 20.2 points at 4B, 9B and 27B, and pass rate by 10--13 points at every size (\Figref{fig:pdb-wild}).
The mean edit shrinks from 37.8 to 10.8 lines at 4B, from 33.9 to 12.0 at 9B and from 18.4 to 6.8 at 27B.
\pc{} outperforms EA-GRPO across all sizes in both unit test pass rate and precision.
On \pdb{}-Wild, unit-test-only GRPO shows no improvement in precision either, changing it by $-2.7$, $+0.1$ and $-9.0$ points against the base model at 4B, 9B and 27B.

\paragraph{\pc{} resolves more issues with smaller patches on SWE-Bench Verified.}
With SWE-agent~\citep{sweagent} as the harness (\Figref{fig:swebench-gt}), \pc{}-9B resolves 32.7\% of issues~\citep{swebenchverified}, against 25.9\% for the base model, 28.1\% for unit-test-only GRPO and 27.1\% for EA-GRPO; at 27B, \pc{} resolves 70.8\% against 67.9\% for the base model and 69.5\% for unit-test-only GRPO.
Measured as the mean excess of changed lines over the gold patch on resolved instances, unit-test-only GRPO over-edits by 58\% and EA-GRPO by 37\%, while \pc{}-9B holds the excess at only 29\% (27B: 42\% vs.\ 47\% for GRPO and 62\% for base).
When the gold patch is applied before the agent starts, \pc{}-27B breaks only 8.9\% of repositories against the base model's 11.0\%, while GRPO raises it to 13.1\%.
Taken together, \pc{} is the only method that improves resolution rate without inflating patch size or damaging correct code.

\paragraph{Precision transfers to multi-turn agentic editing, and agentic training closes the pass-rate gap.}
In a five-turn agent with \texttt{view}, \texttt{str\_replace\_editor} and \texttt{submit} tools on \pdb{} Test (\Figref{fig:agentic}), the single-turn-trained \pc{}-9B retains 52.6\% edit precision, against 42.9\% for unit-test-only GRPO, but trails it by 2.9 points in pass rate.
Trained instead with multi-turn agentic RL on the same tasks, \pc{}-9B-Agent reaches 67.4\% precision at a 56.3\% pass rate, the best of every 9B model in agentic debugging on both.

\subsection{Weak-to-Strong Transfer}
\label{sec:transfer}

Reasoning can move between models without their weights: weak-to-strong generalization elicits a strong model's capability from a weaker supervisor~\citep{weak2strong,w2sreasoning}, advisor models steer a black-box model with natural-language advice~\citep{advisormodels}, and \citet{reasoningtravels} prefill one model's chain of thought into another.
We ask whether an editing discipline, rather than a capability, travels this way from a small trained model to frontier models that already exceed it on correctness.
A trained \pc{} model, the \emph{donor}, thinks about a task; we prefill its think block, token for token, into the assistant turn of an untrained \emph{receiver} over a raw completions endpoint, and the receiver writes the repair, either closing the think block at once or continuing to think from its end.
No weights, tokenizer or training are shared.

\paragraph{The trace carries both correctness and precision.}
Prefilled into its own untrained base model, which answers without further thinking (\Tabref{tab:transfer}, own base), the \pc{}-9B think block recovers \textbf{73\%} of the pass-rate gain that training produced and \textbf{60\%} of the precision gain, and it cuts the base model's edit from 13.7 to 6.5 lines.
With no weights changed, the think block alone moves the base model most of the way toward the trained model on both axes.

\paragraph{A 9B donor teaches larger and stronger models to edit precisely.}
Handed the same block (\Tabref{tab:transfer}, last two blocks), five receivers that the donor cannot match on correctness each gain 4.6--12.4 points of edit precision while their pass rate moves by $-1.5$ to $+4.3$ points.
As a control, the untrained 9B's block costs the 27B receiver 7.4 points of pass rate for 2.6 of precision.
Three receivers continue thinking from the end of the prefilled block; Qwen3.8-Max~\citep{qwen2026qwen38max} and GPT-5.6 Sol, whose APIs cannot prefill, receive it as text in the user turn.
The largest gains go to Qwen3.8-Max (precision +12.4\%) and Qwen3.6-27B (precision +11.1\%, pass rate +4.3).
Even \textit{GPT-5.6 Sol (high) improves on both pass rate and precision with the donor thinking}, which ends above the donor on correctness and only 1.8\% behind it on precision.

\begin{figure}[t]
\begin{minipage}[t]{0.5547\linewidth}
\vspace{0pt}
\centering
\includegraphics[width=\linewidth]{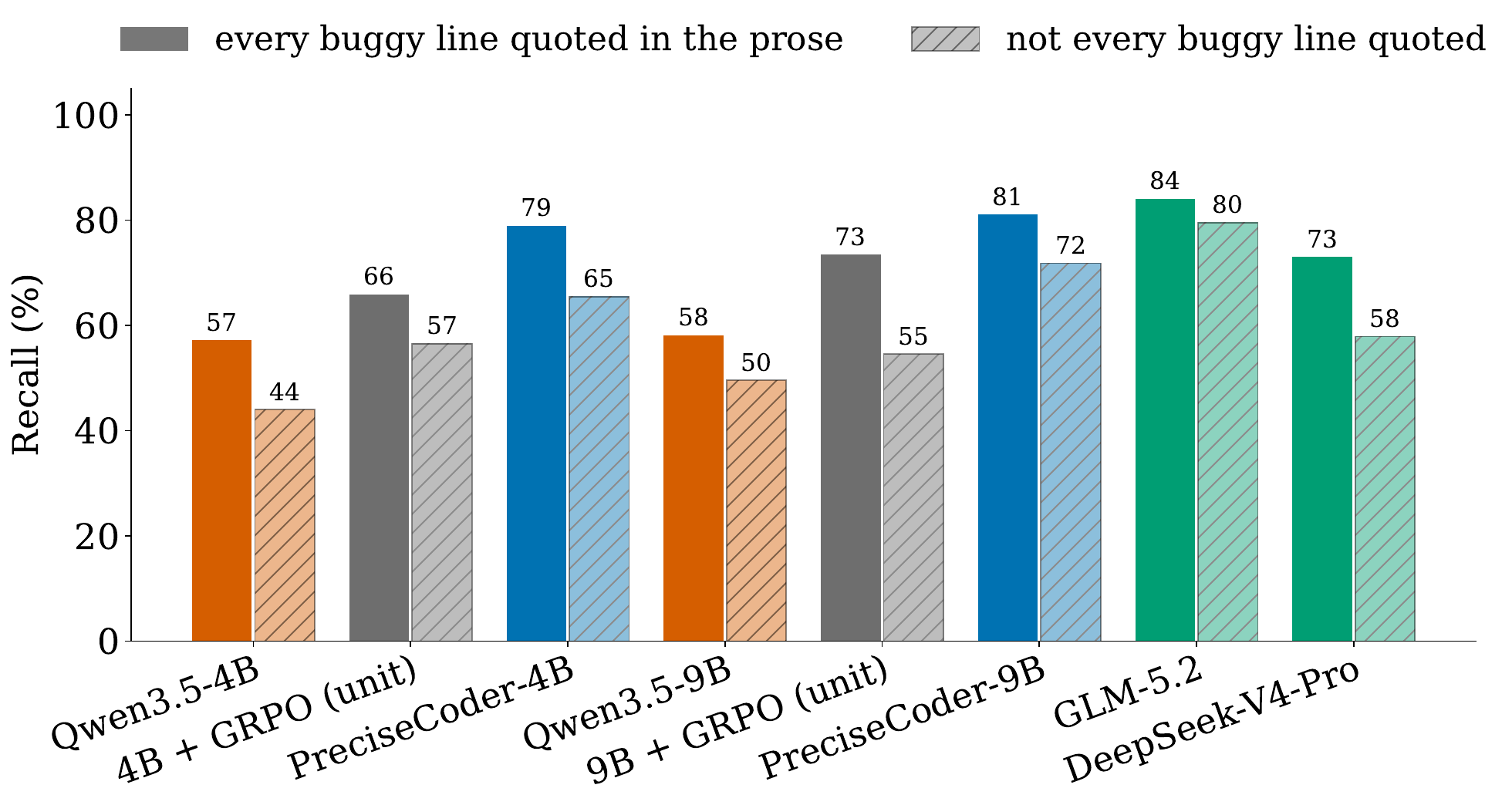}
\caption{Recall on \pdb{} Test when the thinking quotes every buggy line verbatim is consistently higher than when it does not.}
\label{fig:thinking-grounding}
\end{minipage}\hfill
\begin{minipage}[t]{0.4153\linewidth}
\vspace{0pt}
\centering
\includegraphics[width=\linewidth]{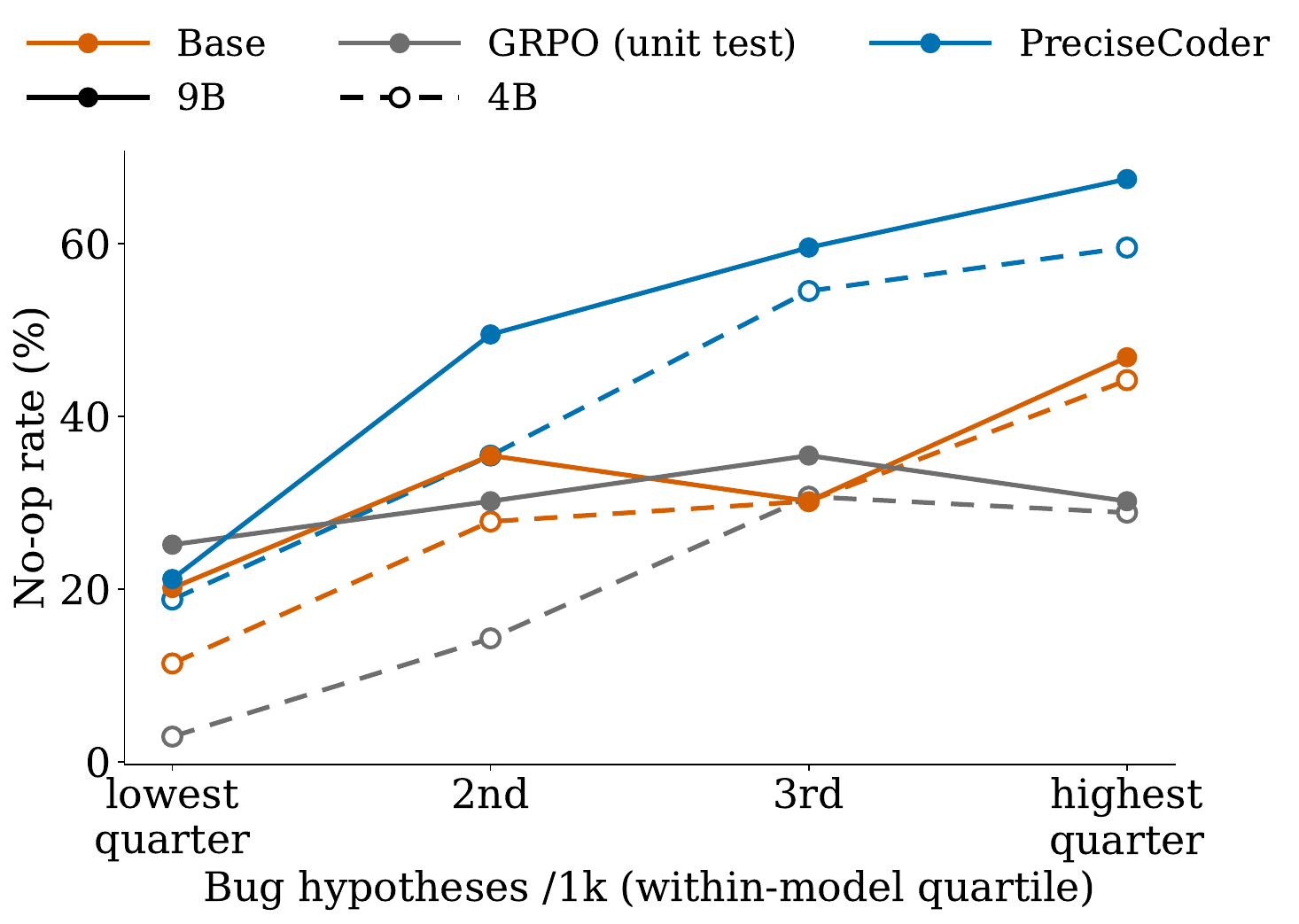}
\caption{GT Debug no-op rate by within-model quartile of bug hypotheses. More hypotheses, more restraint.}
\label{fig:thinking-bughunt}
\end{minipage}
\end{figure}

\subsection{What Changed in the Thinking}
\label{sec:thinking}

To see what the block carries, we analyze the complete chains of thought on \pdb{} Test of the 4B and 9B base, unit-test-only GRPO and \pc{} models, and of two frontier models whose traces are readable, GLM-5.2 and DeepSeek-V4-Pro, on the lexical categories, distinctive bigrams by weighted log-odds, and verbatim matches of the buggy lines (details in Appendix~\ref{app:thinking_transfer}).

\paragraph{RL makes the model think longer, only \pc{} changes what it says about a line.}
Under either reward the 9B thinks to the 4,096-token budget, quotes code inline half again as often (170 and 168 spans per 1,000 words against 111) and backtracks more (``wait'' 12.3 and 15.2 per 1,000 words against 6.1; see Appendix~\Figref{fig:thinking-marker}).
What only the precision reward adds is \textit{judgement}: \pc{}-9B hedges more than unit-test-only GRPO (10.8 against 6.5 per 1,000 words: ``might be'', ``I think'', ``I'll assume''), echoes the instructions half as often (3.4 against 6.5) and almost never argues from the prompt that a bug must exist (0.04 against 0.59).
\pc{}'s distinctive bigrams are ``might be'', ``is valid'' and ``keep the'', weighing whether a line is worth keeping, while GRPO's are ``let's check'', ``only fix'' and ``make minimal'', restating the instruction to fix.

\paragraph{Quoting the buggy line is the strongest linguistic correlate of a correct fix.}
The share of tasks on which the thinking names every buggy line verbatim rises from 36\% to 67\% at 4B and from 43\% to 71\% at 9B under \pc{}.
Within each of the eight models in \Figref{fig:thinking-grounding}, recall is 4--19\% higher on the tasks whose thinking quotes every buggy line.
Holding the model fixed, one standard deviation more of quoting is worth +4.3 $F_1$ and +3.2 precision points. 

\paragraph{The trained model does not come to think like a frontier model.}
Frontier models have different own thinking patterns, which hinder the effectiveness of SFT training before RL.
GLM-5.2 talks to itself in the first person and DeepSeek-V4-Pro thinks in questions, whereas the trained models think in identifier-anchored conditionals (``\emph{x} is'', ``if \emph{x}'', ``\emph{x} returns'' for a quoted identifier \emph{x}; see Appendix~\Tabref{tab:thinking-pdb}).
The receivers of \Tabref{tab:transfer} thus gain precision from a block written in a style unlike their own; what they take from it, we hypothesize, is the line it names and the verdict it reaches on each edit.

\paragraph{On correct code, restraint comes from proposing bugs and rejecting them.}
On the 1,512 correct programs of GT Debug, every model asserts that the code ``looks correct'' at the same rate (8--11 times per 1,000 words); what differs is what it does next.
\pc{}-9B proposes 27.6 bug hypotheses per 1,000 words against 4.7 for its base and 10.8 for unit-test-only GRPO, and its no-op rate rises from 21\% to 68\% between the quarter of tasks with the fewest hypotheses and the quarter with the most, whereas the 9B unit-test-only model stays near 30\%, as shown in \Figref{fig:thinking-bughunt}.
Every trained model edits 85--99\% of the clean programs on which its thinking confidently stops before the 16k-token budget.
The precision reward does not teach the model to declare a program correct early; it teaches it to keep testing candidate bugs and, finding none it trusts, to return the code unchanged.

\section{Discussion}
\label{sec:discussion}

A unit-test reward scores the program a model returns, not the edit that produced it, so RL from unit tests raises pass rates while letting edits grow and teaching the model to invent bugs in correct code.
\pc{} makes the reward see the edit through semantically verified precision and recall: it improves debugging fidelity and keeps unit-test pass rate on-par with standard training.
The learned precision is not tied to the training distribution: it carries to multi-line and repository-level faults, to SWE-bench Verified, where \pc{} resolves more issues with smaller patches and breaks fewer already-fixed repositories.
Most importantly, it travels without weights: the think block of a 9B \pc{} model raises the edit precision of frontier models such as GPT-5.6 Sol.
Since the trained model thinks nothing like its receivers, we attribute the transfer to what the block does rather than how it sounds: it names the buggy line and weighs whether each edit is worth keeping.

The idea is not specific to debugging: the reward credits an action only when it is checked against a reference and verified to do the work the reference requires, so a correct but unnecessary action earns nothing.
In long-horizon agentic tasks, where each unneeded action is a chance to break something, such a reward could serve both performance and safety by penalizing side effects that an outcome-only reward cannot see; the lower damage of \pc{}-27B on already-fixed repositories is a first sign, and defining the reference there is left to future work.

\subsection*{Ethics statement}

This work trains language models to repair bugs in code with minimal edits. It
involves no human subjects, and uses only publicly released datasets and
benchmarks and open-weight models under their respective licenses. All model-generated code is executed in a sandbox during training and
evaluation. Better automated debugging could lower the cost of maintaining
software. Like any code model, it could also be misused or produce subtly wrong
fixes, and precise minimal edits may make such errors harder to spot in review; we
therefore recommend human review of model-proposed patches. We are not aware of
any conflicts of interest.

\subsection*{Reproducibility statement}

We describe the reward design and optimizer in \S\ref{sec:method} and the training
and evaluation setup in \S\ref{sec:exp-setup}. Appendix~\ref{app:implementation}
lists hyperparameters and baseline configurations, Appendix~\ref{app:prompts} gives
every prompt template, and Appendix~\ref{app:ordering} proves the pass/fail
ordering property in full. Data splits are made by base task, so variants of the
same problem never cross splits. We will release code, data splits and checkpoints.

\bibliography{preprint}
\bibliographystyle{preprint}

\appendix
\clearpage

\section*{Appendix}
\startcontents[appendix]
\renewcommand{\thesection}{\Alph{section}} 
\printcontents[appendix]{}{1}{\setcounter{tocdepth}{3}}
\setcounter{section}{0}

\section{More Reward Analysis}
\label{app:reward_analysis}

\subsection{Pass/fail ordering under decoupled normalization}
\label{app:ordering}
Let $p\in(0,1)$ be the fraction of passing completions in a group of size $G$.
With population standardization, a passing completion has $z^u=\sqrt{(1-p)/p}$
and a failing one $z^u=-\sqrt{p/(1-p)}$, a gap of $1/\sqrt{p(1-p)}\ge 2$. For
any $G$ standardized values, the gap between two of them is at most
$\sqrt{2G}$, attained with the two at $\pm d$ and the rest at the mean. A
passing completion therefore has the larger advantage whenever
$(1-\lambda)\cdot 2>\lambda\sqrt{2G}$, i.e.\ $\lambda<b=2/(2+\sqrt{2G})$.
Batch-level rescaling is a positive affine map and preserves the order. This
is a worst-case bound; since $F_1$ and $\mathrm{unit}$ are positively
correlated (\Figref{fig:roles}), the configuration that attains it is rare.

\subsection{Collapsed terms in groups}

\paragraph{$F_1$ supplies an ordering where the unit test is silent.}
$\mathrm{unit}$ is constant on 43--46\% of groups of six base-model samples,
$F_1$ on only 17--24\%. What $F_1$ adds depends on the
group. In uniformly passing groups (51\% of training groups in the \pc{} run
of \figref{fig:training-dynamics}), $\mathrm{unit}$ and recall are identically
one and $F_1$ reduces to an increasing function of precision, so completions
are ranked by the minimality of their edit. In uniformly failing groups
(14\%), recall gives partial credit for the faults that were repaired, so a
repair of one fault out of two outranks a repair of none and the group keeps a
gradient. In mixed groups (35\%), a weight $\lambda<b$ keeps every
passing completion above every failing one and $F_1$ breaks ties within each
side. EA-GRPO's edit-line penalty also restores a gradient in 34\% of
uniformly passing groups, as ours does (33\%), but it multiplies the unit test,
so every uniformly failing group still collapses, as under the unit test alone.
Replaying these rollouts through each reward, the advantage collapses on 65\%
of all groups under the unit-test reward alone and 48\% under EA-GRPO,
compared to 39\% under ours; 33 of those 39 points are groups in which all six
rollouts return the exact same passing fix.

\section{Additional Results}
\label{app:results}

\subsection{GRPO + clean rows}
\label{app:gt_train}

Every \pdb{} training row contains an injected bug, so ``there is something to fix'' is a constant of
the training distribution. GRPO + clean rows adds the missing case and tests whether restraint can be taught directly. To the 5{,}274 buggy training rows we
add 20\% extra rows per dataset (577 BigCodeBench and 478 LiveCodeBench, 1{,}055 in all) whose input is
the reference solution itself, spread evenly over the base problems of the training split; the prompt
is the same minimal-edit prompt, so the model is not told that the code may be correct. Buggy rows
are rewarded with the unit-test reward alone, as in the unit-test-only GRPO control. Clean rows are
rewarded with the sum of two group-standardized terms, the unit-test outcome and the negative number
of changed lines, so the only answer that scores well is to return the code unchanged. Training
otherwise follows the recipe of Section~\ref{sec:exp-setup}. Because these runs see bug-free inputs during training,
their GT Debug rates are marked with an asterisk.

\begin{table}[!htb]
\caption{GRPO + clean rows against the base model, unit-test-only GRPO and \pc{} (the latter three as in \Tabref{tab:main}).
$^*$Trained on bug-free inputs. Bold: best within each size.}
\label{tab:gt-train}
\centering
\small
\setlength{\tabcolsep}{4pt}
\begin{tabular}{lcccccc}
\toprule
& \multicolumn{4}{c}{\pdb{} Test} & \multicolumn{2}{c}{GT Debug} \\
\cmidrule(lr){2-5} \cmidrule(lr){6-7}
Model & Pass (\%)\,$\uparrow$ & Prec. (\%)\,$\uparrow$ & Recall (\%)\,$\uparrow$ & Lines\,$\downarrow$ & No-op (\%)\,$\uparrow$ & Damage (\%)\,$\downarrow$ \\
\midrule
Qwen3.5-4B            & 36.5 & 35.8 & 48.5 & 15.3 & 30.6 & 19.3 \\
\quad + GRPO          & 55.1 & 30.0 & 60.1 & 13.8 & 19.3 & 17.8 \\
\quad + GRPO + clean rows & 50.6 & 53.9 & 66.0 & 3.8 & \textbf{73.7}$^*$ & \textbf{7.9}$^*$ \\
\pc{}-4B              & \textbf{55.3}\,$\pm$\,0.9 & \textbf{64.9}\,$\pm$\,1.3 & \textbf{72.9}\,$\pm$\,1.1 & \textbf{3.0}\,$\pm$\,0.1 & 48.0\,$\pm$\,5.8 & 13.5\,$\pm$\,1.5 \\
\midrule
Qwen3.5-9B            & 41.2 & 43.9 & 53.1 & 13.7 & 36.6 & 18.7 \\
\quad + GRPO          & \textbf{63.9} & 41.7 & 68.4 & 11.9 & 30.7 & 11.4 \\
\quad + GRPO + clean rows & 61.3 & 55.1 & 71.1 & 5.9 & \textbf{71.8}$^*$ & \textbf{7.9}$^*$ \\
\pc{}-9B              & 62.5\,$\pm$\,1.9 & \textbf{67.7}\,$\pm$\,1.5 & \textbf{76.8}\,$\pm$\,1.1 & \textbf{3.0}\,$\pm$\,0.1 & 51.6\,$\pm$\,2.7 & 11.8\,$\pm$\,0.9 \\
\bottomrule
\end{tabular}
\end{table}

\Tabref{tab:gt-train} compares the two runs with the base model, unit-test-only GRPO and \pc{}.
Training on bug-free code gives the highest no-op rate and the lowest damage of any model of its size
(73.7\% and 71.8\% no-op, 7.9\% damage at both sizes, against 48.0\% and 51.6\% no-op for \pc{}), but it
gives up pass rate when the input is genuinely buggy: 50.6\% against 55.1\% for unit-test-only GRPO at 4B
and 61.3\% against 63.9\% at 9B. The buggy rows of these runs carry no edit penalty, only the unit-test
reward of the GRPO control; the clean rows alone still raise precision over unit-test-only GRPO by 23.9
points at 4B and 13.4 at 9B and cut the edit by 10.0 and 6.0 lines. Learning to leave correct code alone
thus also teaches the model to localize its fixes, which further indicates that precise debugging and bug
hallucination are one problem that can be solved jointly.

\FloatBarrier

\subsection{Out-of-Distribution Generalization}
\label{app:generalization}

Training on precise repairs could narrow what the model can do outside the
repair task. \Tabref{tab:generalization} checks this on five benchmarks the
model never trains on: from-scratch code generation on BigCodeBench~\citep{bigcodebench} and
LiveCodeBench~\citep{livecodebench} problems that appear in no \pdb{} split, HumanEval+~\citep{evalplus}, CruxEval~\citep{cruxeval} and
DebugBench~\citep{debugbench}. At both sizes \pc{} scores above its base model on all five. At
4B it is also above unit-test-only GRPO on every benchmark except DebugBench,
while GRPO loses 21.3 points on CruxEval. At 9B, GRPO is ahead on four of the
five by at most 1.5 points, and \pc{}-9B is ahead on LiveCodeBench by 5.3
points.

\begin{table}[!htb]
\caption{Out-of-distribution generalization: unit-test pass rate (\%), one sample per task.
\pc{}-9B is the mean of the three Table~\ref{tab:main} seeds.}
\label{tab:generalization}
\centering
\small
\begin{tabular}{lccccc}
\toprule
& \multicolumn{2}{c}{From scratch} & & & \\
\cmidrule(lr){2-3}
Model & BCB (901) & LCB (806) & HumanEval+ & CruxEval & DebugBench \\
\midrule
\multicolumn{6}{l}{\textit{4B}} \\
Qwen3.5-4B (base)       & 43.2 & 42.4 & 86.6 & 85.9 & 67.0 \\
\quad + GRPO            & 44.2 & 49.8 & 86.6 & 64.6 & \textbf{75.5} \\
PreciseCoder-4B (ours)  & \textbf{46.5} & \textbf{53.3} & \textbf{92.1} & \textbf{90.5} & 72.9 \\
\midrule
\multicolumn{6}{l}{\textit{9B}} \\
Qwen3.5-9B (base)       & 45.7 & 45.2 & 90.2 & 89.2 & 75.3 \\
\quad + GRPO            & \textbf{48.4} & 52.6 & \textbf{92.7} & \textbf{91.6} & \textbf{78.6} \\
PreciseCoder-9B (ours)  & 47.6 & \textbf{57.9} & 91.7 & 90.7 & 77.1 \\
\bottomrule
\end{tabular}
\end{table}

\FloatBarrier
\subsection{Reward-Weight Search}
\label{app:k-search}

\begin{figure}[!htb]
\centering
\includegraphics[width=0.55\linewidth]{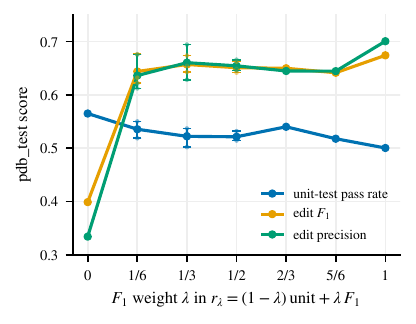}
\caption{Reward-weight sweep on Qwen3.5-4B GRPO, \pdb{} Test. Lines are seed means; whiskers span
three seeds at $\lambda\in\{1/6,1/3,1/2\}$.}
\label{fig:k-search}
\end{figure}

\Figref{fig:k-search} sweeps the $F_1$ weight $\lambda$ in
$r_\lambda=(1-\lambda)\,\mathrm{unit}+\lambda\,F_1$ for Qwen3.5-4B under GRPO. Any $\lambda>0$ changes
the model's behavior: from $\lambda=0$ to $\lambda=1/6$ edit precision rises from 33.4\%
to 63.6\% and $F_1$ from 39.9\% to 64.4\%, at a cost of 2.9 pass-rate points.
Across $\lambda\in[1/6,5/6]$ each of the three metrics varies by at most 2.5
points, no more than the spread between seeds at a single $\lambda$. We
use $\lambda=1/3$.

\subsection{Prompting analysis on GT Debug Task}
\label{app:prompt_gt_debug}

\citet{fixedbench} show that an abstain-or-fix prompt reduces edits when a coding agent is given already-correct code.
We test such a prompt on GT Debug (\Tabref{tab:gt-prompt}, Appendix~\ref{app:prompt-abstain}): it asks the model to check whether the code has a bug and to return it unchanged if not.
The prompt helps every model: each raises its no-op rate by 14 to 44 points and lowers its damage rate.
Base models gain the most (34--44 points of no-op), with frontier models close behind (27--34).
Unit-test-only GRPO remains the least restrained model at 4B and 9B (51.4\% and 51.5\% no-op), and at 27B it falls below its own base model (67.9\% vs.\ 75.5\%).
\pc{} beats the base model and EA-GRPO on both rates at 4B and 9B, beats frontier models at 27B, and narrows the gap to runs trained on clean rows: from 25.7 to 11.3 no-op points at 4B and from 20.2 to 4.3 at 9B, where its damage (7.4\%) falls below theirs (8.0\%).

\begin{table}
\centering
\captionof{table}{GT Debug under the abstain-or-fix prompt; $\Delta$ = change from the minimal prompt.}
\label{tab:gt-prompt}
\footnotesize
\setlength{\tabcolsep}{2pt}
\begin{tabular}{lrrrr}
\toprule
& \multicolumn{2}{c}{Abstain-or-fix} & \multicolumn{2}{c}{$\Delta$} \\
\cmidrule(lr){2-3} \cmidrule(lr){4-5}
Model & No-op & Dmg. & No-op & Dmg. \\
\midrule
Qwen3.5-4B          & 68.3 & 11.5 & $+$37.7 & $-$7.8 \\
GRPO-4B             & 51.4 & 10.9 & $+$32.2 & $-$6.9 \\
EA-GRPO-4B          & 57.0 & 14.9 & $+$13.7 & $-$3.9 \\
PreciseCoder-4B     & \bf 71.9 & \bf 7.6 & $+$23.8 & $-$5.9 \\
\midrule
Qwen3.5-9B          & 71.0 & 9.6 & $+$34.4 & $-$9.2 \\
GRPO-9B             & 51.5 & 10.1 & $+$20.8 & $-$1.3 \\
EA-GRPO-9B          & 67.7 & 8.9 & $+$20.7 & $-$4.6 \\
PreciseCoder-9B     & \bf 73.7 & \bf 7.4 & $+$22.1 & $-$4.4 \\
\midrule
Qwen3.6-27B         & 75.5 & 7.1 & $+$44.4 & $-$8.4 \\
GRPO-27B            & 67.9 & 7.3 & $+$41.0 & $-$5.3 \\
PreciseCoder-27B    & \bf 80.4 & \bf 5.8 & $+$33.1 & $-$4.5 \\
\midrule
DeepSeek-V4.1-Flash & 74.0 & 10.0 & $+$34.4 & $-$11.5 \\
Qwen3.8-Flash-Next  & 48.7 & 14.6 & $+$26.6 & $-$3.3 \\
\bottomrule
\end{tabular}
\end{table}

\FloatBarrier
\subsection{Chain-of-Thought Transfer from Every 9B Donor}
\label{app:cot-transfer-27b}

\Secref{sec:transfer} hands only the \pc{}-9B think block to Qwen3.6-27B.
\Tabref{tab:cot-transfer-27b} repeats that transfer with every 9B model of
\Tabref{tab:main} as the donor, under the same protocol. The precision the
27B receiver gains grows with how precisely the donor was trained to edit:
$+2.6$ points from the untrained 9B base, $+7.4$ from unit-test-only GRPO,
$+10.1$ on average from the three \pc{}-9B seeds and $+11.1$ from EA-GRPO.
Only the trained donors keep or raise the receiver's pass rate; the base
model's trace costs 7.4 points. The \pc{}-9B traces give the largest pass-rate gain,
$+2.8$ points on average and $+4.3$ for seed 42, at EA-GRPO's level of
precision. The seed-42 row is the Qwen3.6-27B entry of \Tabref{tab:transfer}.

\begin{table}[!htb]
\caption{Chain-of-thought transfer to Qwen3.6-27B from every 9B donor, \pdb{} Test. $\Delta$ = change
from the receiver's own reasoning; bold = positive $\Delta$.}
\label{tab:cot-transfer-27b}
\centering
\footnotesize
\setlength{\tabcolsep}{4pt}
\begin{tabular}{lccccc rr}
\toprule
& \multicolumn{5}{c}{\pdb{} Test} & \multicolumn{2}{c}{$\Delta$} \\
\cmidrule(lr){2-6} \cmidrule(lr){7-8}
Receiver input & Pass & Prec. & Rec. & $F_1$ & Lines & Pass & Prec. \\
\midrule
Own reasoning (20k)             & 73.8 & 51.0 & 72.6 & 56.4 & 6.7 & ---  & ---  \\
\midrule
\multicolumn{8}{l}{\textit{+ think block of the 9B donor}} \\
Qwen3.5-9B (base)               & 66.4 & 53.6 & 70.0 & 56.6 & 7.6 & $-$7.4 & \textbf{\boldmath$+$2.6} \\
\quad + GRPO                    & 74.2 & 58.4 & 76.6 & 62.7 & 5.5 & \textbf{\boldmath$+$0.4} & \textbf{\boldmath$+$7.4} \\
\quad + EA-GRPO                 & 76.1 & 62.1 & 77.7 & 65.9 & 5.9 & \textbf{\boldmath$+$2.3} & \textbf{\boldmath$+$11.1} \\
\pc{}-9B, seed 42               & 78.1 & 62.1 & 78.3 & 66.3 & 5.7 & \textbf{\boldmath$+$4.3} & \textbf{\boldmath$+$11.1} \\
\pc{}-9B, seed 43               & 76.4 & 61.0 & 78.2 & 65.1 & 5.4 & \textbf{\boldmath$+$2.6} & \textbf{\boldmath$+$10.0} \\
\pc{}-9B, seed 44               & 75.3 & 60.3 & 77.5 & 64.5 & 5.8 & \textbf{\boldmath$+$1.5} & \textbf{\boldmath$+$9.3} \\
\pc{}-9B, mean                  & 76.6{\tiny$\pm$1.4} & 61.1{\tiny$\pm$0.9} & 78.0{\tiny$\pm$0.4} & 65.3{\tiny$\pm$0.9} & 5.7{\tiny$\pm$0.2}
                                & \textbf{\boldmath$+$2.8} & \textbf{\boldmath$+$10.1} \\
\bottomrule
\end{tabular}
\end{table}

\subsection{Detailed Linguistic Analysis on the Thinking}
\label{app:thinking_transfer}

\paragraph{Setup.}
We analyze the complete think blocks of the six local models of \Tabref{tab:main} on the 1,205 \pdb{} Test tasks and the 1,512 GT Debug programs, and of GLM-5.2 and DeepSeek-V4-Pro on \pdb{} Test, one sample per task at temperature 0.6 under the minimal-edit prompt.
The local models think for at most 4,096 tokens on \pdb{} Test and 16,384 on GT Debug and the frontier APIs for 16k--32k, so every lexical measure is a rate per 1,000 \emph{prose} words, the block with fenced code removed and inline code spans replaced by a placeholder, and length is reported on its own.

We count 15 hand-built, case-insensitive lexical categories, among them backtracking (``wait'', ``actually''), hedging (``might be'', ``I think'', ``I'll assume''), verification, minimal-edit talk, instruction echo, ``looks correct'' assertions, ``the prompt says there is a bug'' and bug hypotheses (``maybe the bug is'', ``I found it''), plus lexical diversity (moving-average type--token ratio,~\citealp{mattr}) and repetition.
Distinctive bigrams are ranked by the weighted log-odds ratio with an informative Dirichlet prior~\citep{fightinwords}, and vocabularies are compared by the Jensen--Shannon divergence of their unigram distributions.

Grounding asks whether every ground-truth buggy line appears verbatim, after whitespace normalization, in the prose, and how early.
Links to performance are Spearman correlations within each model, paired differences on the same task within a size, and ordinary least squares of the metric on the standardized features with model fixed effects and task-clustered standard errors; they are associations, not causal effects.
The lexicons were fixed before any correlation was computed.

\begin{table}[!htb]
\caption{Thinking on \pdb{} Test: length, lexical rates per 1,000 prose words, and grounding in the buggy line,
for the base, unit-test-only GRPO and \pc{} models of \Tabref{tab:main} (\pc{} seed 42, one sample per task) and
two frontier models. Pass and precision are those of this single sample. DeepSeek-V4-Pro here is DeepSeek-V4-Pro Preview, against DeepSeek-V4-Pro (0813) in \Tabref{tab:main}}
\label{tab:thinking-pdb}
\centering
\small
\setlength{\tabcolsep}{4pt}
\resizebox{\linewidth}{!}{
\begin{tabular}{@{}lrrrrrrrr@{}}
\toprule
& \multicolumn{3}{c}{Qwen3.5-4B} & \multicolumn{3}{c}{Qwen3.5-9B} & \multicolumn{2}{c}{Frontier} \\
\cmidrule(lr){2-4} \cmidrule(lr){5-7} \cmidrule(lr){8-9}
& Base & + GRPO & \pc{} & Base & + GRPO & \pc{} & GLM-5.2 & DeepSeek-V4-Pro \\
\midrule
Pass (\%)                          & 36.5 & 55.1 & 56.2 & 41.2 & 63.9 & 64.6 & 70.8 & 67.6 \\
Precision (\%)                     & 35.8 & 30.0 & 64.6 & 43.9 & 41.7 & 67.9 & 65.0 & 40.8 \\
\midrule
Think tokens (mean)                & 2535 & 3617 & 4096 & 2469 & 4096 & 4087 & 7375 & 12043 \\
At the thinking budget (\%)        & 50 & 79 & 100 & 50 & 100 & 100 & 7 & 16 \\
Prose words                        & 932 & 1179 & 1432 & 908 & 1494 & 1581 & 2919 & 4975 \\
Code share of the block (\%)       & 14 & 21 & 8 & 16 & 8 & 4 & 19 & 12 \\
Inline code spans /1k              & 114 & 143 & 155 & 111 & 170 & 168 & 63 & 84 \\
\midrule
Backtracking /1k                   & 11.1 & 19.0 & 15.1 & 10.1 & 14.3 & 20.0 & 18.0 & 11.1 \\
Hedging /1k                        & 4.8 & 4.8 & 5.9 & 5.4 & 6.5 & 10.8 & 7.0 & 9.3 \\
Verification /1k                   & 8.1 & 11.4 & 7.4 & 8.5 & 9.0 & 7.0 & 10.2 & 6.0 \\
Minimal-edit talk /1k              & 3.1 & 2.2 & 2.9 & 5.3 & 4.8 & 3.0 & 7.5 & 3.5 \\
Instruction echo /1k               & 3.7 & 1.6 & 5.2 & 5.5 & 6.5 & 3.4 & 6.1 & 2.8 \\
``Prompt says there is a bug'' /1k & 0.20 & 0.15 & 0.03 & 0.25 & 0.59 & 0.04 & 0.45 & 0.44 \\
``Looks correct'' /1k              & 4.1 & 4.1 & 5.0 & 4.3 & 4.6 & 4.7 & 5.1 & 3.3 \\
Questions /1k                      & 7.8 & 4.0 & 20.0 & 4.7 & 11.5 & 11.4 & 6.5 & 50.8 \\
First person /1k                   & 16.9 & 17.7 & 9.6 & 19.5 & 13.1 & 16.4 & 36.2 & 26.1 \\
\midrule
Every buggy line quoted (\%)       & 36 & 40 & 67 & 43 & 78 & 71 & 61 & 72 \\
Fixed line quoted (\%)             & 22 & 26 & 45 & 27 & 54 & 46 & 55 & 58 \\
First mention (relative position)  & 0.26 & 0.27 & 0.23 & 0.28 & 0.19 & 0.20 & 0.14 & 0.16 \\
\bottomrule
\end{tabular}
}
\end{table}

\begin{figure}[!htb]
\centering
\includegraphics[width=0.72\linewidth]{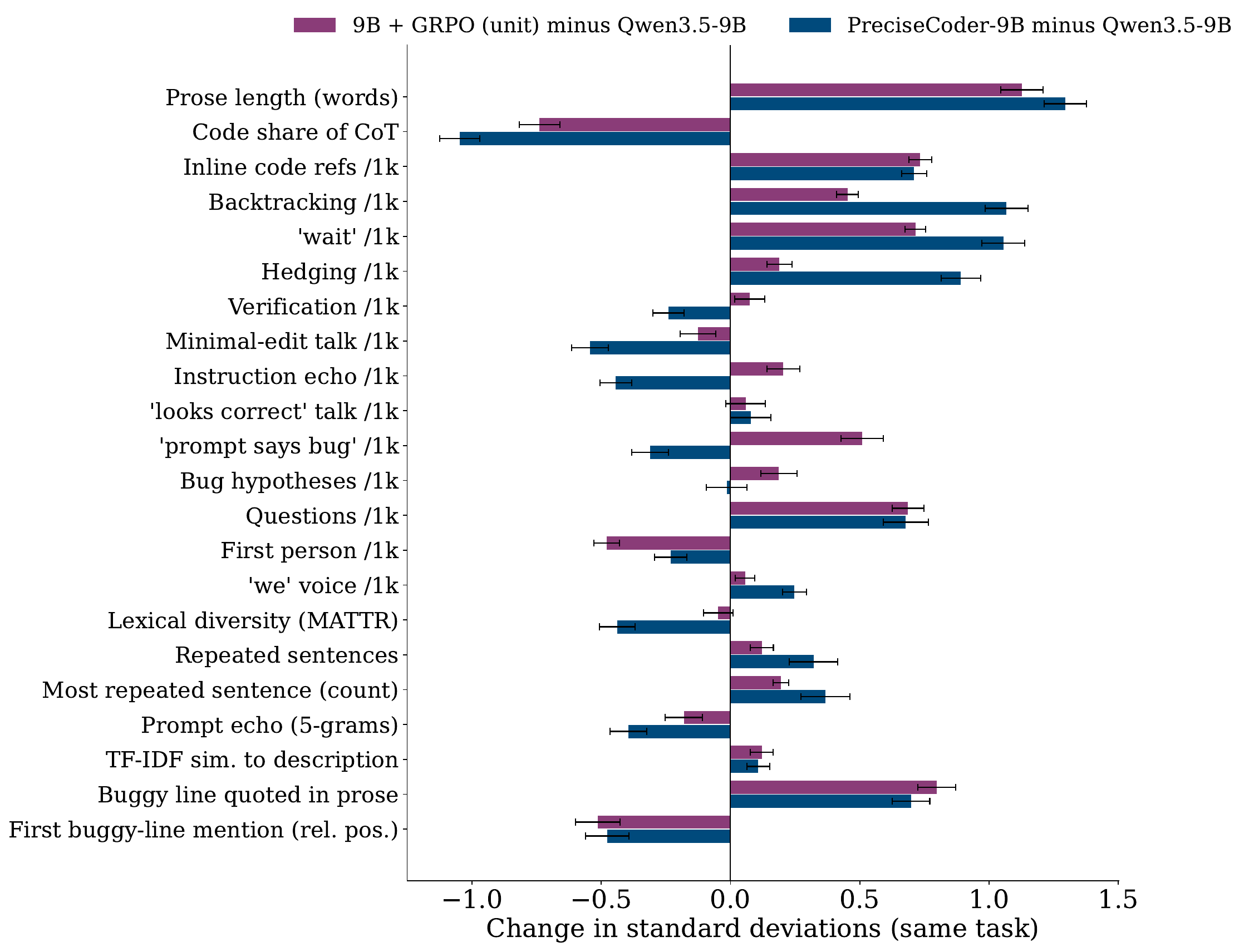}
\caption{What each training changed in the 9B's thinking on the same \pdb{} Test task: unit-test-only GRPO minus
base (purple) and \pc{} minus base (blue), in standard deviations of the feature within the 9B triple; bars are 95\%
confidence intervals.}
\label{fig:thinking-paired}
\end{figure}

\paragraph{Buggy input.}
\Tabref{tab:thinking-pdb} lists the rates behind \Secref{sec:thinking}, and \Figref{fig:thinking-paired} the change each training made on the same task at 9B.
Both rewards produce the same length (every RL model thinks to the budget on 79--100\% of tasks), the same rise in inline code and the same drop in pasted code; on top of that the precision reward adds hedging ($+0.9$ vs.\ $+0.2$ standard deviations), removes minimal-edit talk and instruction echo ($-0.5$ and $-0.4$ vs.\ about $0$ and $+0.2$) and removes ``the prompt says there is a bug'' ($-0.3$ vs.\ $+0.5$).
\Figref{fig:thinking-grounding} shows that recall is higher in every model on the tasks whose thinking quotes every buggy line, by 4 (GLM-5.2) to 19 (9B + GRPO) points.
Surface alignment with the problem statement does not separate the models: the TF-IDF cosine between prose and description is 0.39--0.41 for all six local models, and 2--3\% of prose 5-grams restate the prompt everywhere.

The frontier models write in idiolects no trained model acquires: GLM-5.2 says ``let me'' 17 times per 1,000 words and speaks in the first person 36 times, DeepSeek-V4-Pro asks 51 questions per 1,000 words and repeats its most frequent sentence 14 times per chain, and the Jensen--Shannon divergence to GLM-5.2 grows under either reward (9B: 0.096 for the base, 0.129 for GRPO, 0.127 for \pc{}).

\begin{table}[!htb]
\caption{Thinking on GT Debug (1,512 correct programs, one sample per task, all rows): outcome, length, and the marks
of a bug hunt with nothing to find. Rates per 1,000 prose words unless noted.}
\label{tab:thinking-gt}
\centering
\small
\setlength{\tabcolsep}{5pt}
\resizebox{0.9\linewidth}{!}{
\begin{tabular}{@{}lrrrrrr@{}}
\toprule
& \multicolumn{3}{c}{Qwen3.5-4B} & \multicolumn{3}{c}{Qwen3.5-9B} \\
\cmidrule(lr){2-4} \cmidrule(lr){5-7}
& Base & + GRPO & \pc{} & Base & + GRPO & \pc{} \\
\midrule
No-op (\%)                              & 30.6 & 19.3 & 48.0 & 36.6 & 30.7 & 51.6 \\
Damage (\%)                             & 19.3 & 17.8 & 13.5 & 18.7 & 11.4 & 11.8 \\
Edited lines when edited                & 3.8 & 5.2 & 1.6 & 3.8 & 2.6 & 1.1 \\
\midrule
Think tokens (mean)                     & 9773 & 14102 & 14813 & 9234 & 15868 & 16336 \\
At the thinking budget (\%)             & 54 & 77 & 68 & 51 & 91 & 99 \\
No-op at the budget / below it (\%)     & 46 / 12 & 22 / 2 & 63 / 15 & 55 / 18 & 34 / 1 & 52 / 14 \\
\midrule
Backtracking /1k                        & 23.9 & 38.7 & 30.5 & 21.1 & 43.5 & 85.8 \\
Hedging /1k                             & 14.7 & 15.4 & 13.1 & 14.9 & 10.0 & 51.1 \\
``Looks correct'' /1k                   & 8.2 & 8.0 & 10.1 & 8.1 & 10.9 & 10.2 \\
Bug hypotheses /1k                      & 6.0 & 3.7 & 10.3 & 4.7 & 10.8 & 27.6 \\
``I found it'' (per chain)              & 18.8 & 5.2 & 0.3 & 9.7 & 10.6 & 35.4 \\
``Prompt says there is a bug'' /1k      & 1.03 & 0.61 & 0.36 & 1.10 & 0.49 & 0.17 \\
Repeated sentences (\%)                 & 28.7 & 41.1 & 20.5 & 20.4 & 31.5 & 51.3 \\
Most repeated sentence (count)          & 53 & 64 & 72 & 37 & 84 & 195 \\
\bottomrule
\end{tabular}
}
\end{table}

\paragraph{Correct input.}
\Tabref{tab:thinking-gt} reports the same local models on GT Debug, where there is nothing to find.
With 16k tokens available the trained models fill the budget (9B + GRPO on 91\% of tasks, \pc{}-9B on 99\%), and their chains become a bug hunt: \pc{}-9B backtracks 86 times per 1,000 words, proposes 28 bug hypotheses, writes ``I found it'' 35 times per chain and repeats its most frequent sentence 195 times.
The unit-test-only 9B shows the same pattern at half the intensity, and ``looks correct'' assertions are equally common in every model.
Restraint arrives at the budget: every local model leaves clean code alone far more often when its thinking hits the cap than when it stops early (\pc{}-9B 52\% vs.\ 14\%, 9B + GRPO 34\% vs.\ 1\%), because a chain that stops early has ``found'' a bug.

For the precision-trained models the no-op rate rises with the number of hypotheses the chain raises and rejects (\figref{fig:thinking-bughunt}), from 21\% to 68\% between the quarter of tasks with the fewest and the quarter with the most for \pc{}-9B, whereas the 9B unit-test-only model stays near 30\%.
\pc{}-4B is the exception among the trained models: it loops least (20\% repeated sentences) and instead returns an empty answer on 6.7\% of programs when its thinking exhausts the answer budget.

\begin{figure}[!htb]
\centering
\includegraphics[width=0.72\linewidth]{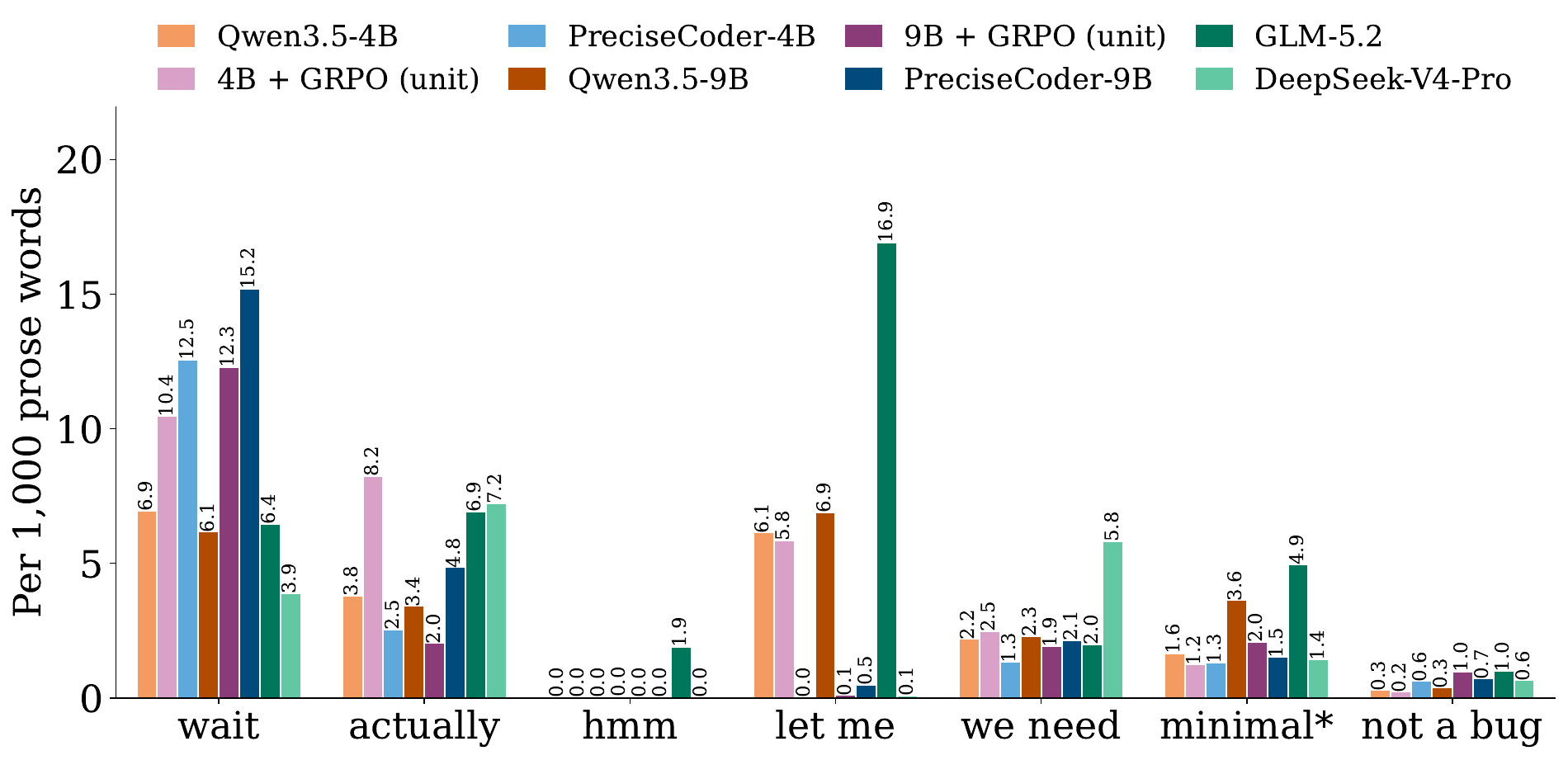}
\caption{Common word markers in the thinking traces.}
\label{fig:thinking-marker}
\end{figure}

\subsection{Result Visualizations}
\label{app:visualizations}

\begin{figure}[!htb]
\centering
\includegraphics[width=0.55\linewidth]{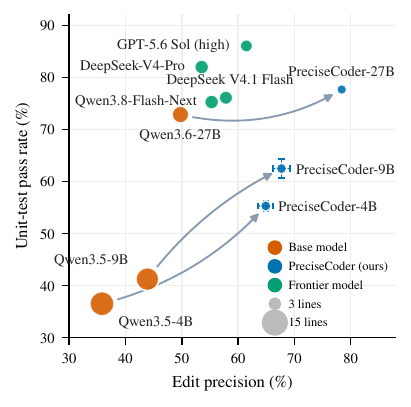}
\caption{PDB Test edit precision against pass rate (bubble area = edited lines). Arrows join each base model to
its \pc{} model; error bars are $\pm1$ sd over three runs.}
\label{fig:main-bubble}
\end{figure}

\begin{figure}[!htb]
\centering
\includegraphics[width=0.8\linewidth]{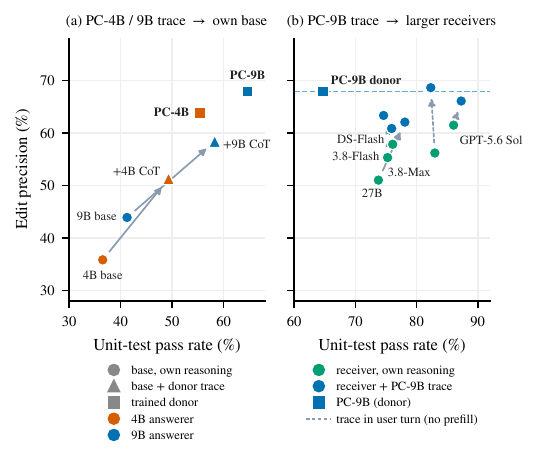}
\caption{Chain-of-thought transfer on PDB Test. (a) A \pc{} think block prefilled into its own base model.
(b) The \pc{}-9B think block handed to larger receivers; dashed arrows mark the trace given in the user turn.}
\label{fig:cot-transfer}
\end{figure}

\begin{figure}[!htb]
\centering
\includegraphics[width=\linewidth]{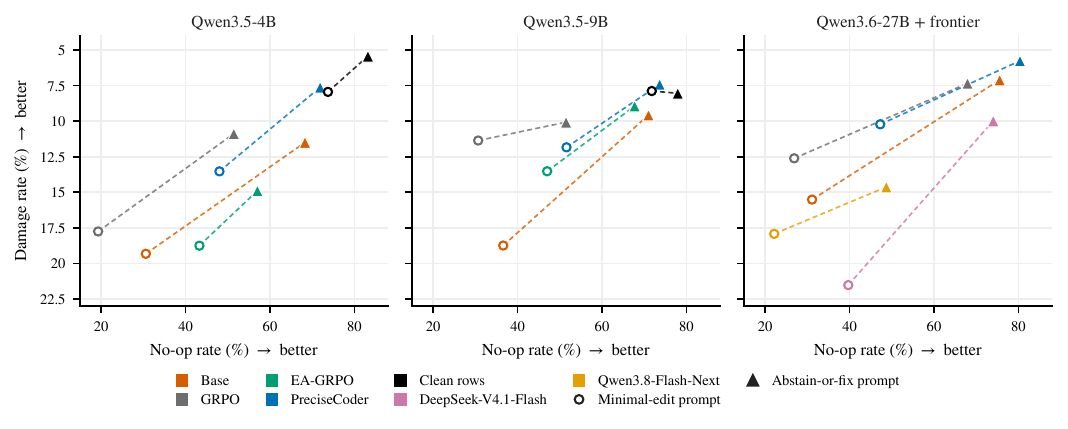}
\caption{GT Debug under the minimal-edit (circles) and abstain-or-fix (triangles) prompts. Up and right is
better.}
\label{fig:gt-prompt-all}
\end{figure}

\Figref{fig:main-bubble} plots the PDB Test columns of \Tabref{tab:main} for the base,
\pc{} and frontier models, \figref{fig:cot-transfer} the chain-of-thought transfer of
\Tabref{tab:transfer}, and \figref{fig:gt-prompt-all} the prompt ablation of
\Tabref{tab:gt-prompt}.

\FloatBarrier
\section{Implementation Details}
\label{app:implementation}

\subsection{Hyperparameters}
\label{app:hyperparams}

\begin{table}[t]
\centering
\caption{Training and evaluation hyperparameters. The upper block is shared by every
reinforcement-learning run, including the GRPO and EA-GRPO controls; the middle block lists
what changes with model size. In the CoT-transfer block, ``trace + $n$'' means the donor's
trace is prefilled into the receiver's thinking, which may continue for $n$ more tokens before it is
closed. Settings not listed are verl defaults.}
\label{tab:hyperparams}
\small
\begin{tabular}{@{}lccc@{}}
\toprule
& \multicolumn{3}{c}{Value} \\
\midrule
\multicolumn{4}{@{}l}{\emph{Reinforcement learning (all sizes)}} \\
Advantage estimator & \multicolumn{3}{c}{GDPO, $\lambda=\tfrac13$ (\S\ref{sec:method:optimizer})} \\
Group size $G$ & \multicolumn{3}{c}{6} \\
Prompts per step / PPO mini-batch & \multicolumn{3}{c}{32 / 16} \\
PPO epochs per step & \multicolumn{3}{c}{1} \\
Optimizer & \multicolumn{3}{c}{AdamW ($\beta_1{=}0.9$, $\beta_2{=}0.999$, weight decay 0.01)} \\
Learning rate & \multicolumn{3}{c}{$1\times10^{-6}$, constant, no warmup} \\
Gradient-norm clip & \multicolumn{3}{c}{1.0} \\
Clip ratio (low / high) & \multicolumn{3}{c}{0.2 / 0.28} \\
KL loss (low-variance estimator) & \multicolumn{3}{c}{0.005} \\
Entropy bonus & \multicolumn{3}{c}{0} \\
Rollout temperature / top-$p$ & \multicolumn{3}{c}{1.0 / 1.0} \\
Max prompt / response tokens & \multicolumn{3}{c}{4{,}096 / 6{,}144} \\
Thinking budget & \multicolumn{3}{c}{4{,}096 tokens} \\
Training epochs & \multicolumn{3}{c}{2} \\
Validation and checkpoint interval & \multicolumn{3}{c}{25 steps} \\
\midrule
& 4B & 9B & 27B \\
\midrule
Base model & Qwen3.5-4B & Qwen3.5-9B & Qwen3.6-27B \\
GPUs & 4$\times$GH200 & 4$\times$H200 & 8$\times$GH200 \\
Rollout tensor parallelism & 1 & 1 & 2 \\
Reported checkpoint (step) & 328 & 325 & 328 \\
Seeds & 3 & 3 & 1 \\
\midrule
\multicolumn{4}{@{}l}{\emph{Evaluation (all models)}} \\
Temperature / top-$p$ / top-$k$ & \multicolumn{3}{c}{0.6 / 0.95 / 10} \\
Samples per task & \multicolumn{3}{c}{1} \\
Thinking budget / max tokens, \pdb{} & \multicolumn{3}{c}{4{,}096 / 6{,}144} \\
Thinking budget / max tokens, GT Debug & \multicolumn{3}{c}{16{,}384 / 20{,}480} \\
\midrule
\multicolumn{4}{@{}l}{\emph{CoT transfer (\S\ref{sec:transfer}): thinking / answer tokens, own reasoning vs.\ + \pc{}-9B trace}} \\
Donor trace (\pc{}-9B) & \multicolumn{3}{c}{4{,}096 (its \pdb{} thinking budget)} \\
Qwen3.5-9B & \multicolumn{3}{c}{4{,}096 / 2{,}048 \quad vs.\quad trace + 0 / 2{,}048} \\
Qwen3.6-27B & \multicolumn{3}{c}{20{,}480 / 2{,}048 \quad vs.\quad trace + 16{,}000 / 2{,}048} \\
Qwen3.8-Flash-Next, DeepSeek V4.1 Flash & \multicolumn{3}{c}{20{,}480 / 2{,}048 \quad vs.\quad trace + 16{,}384 / 2{,}048} \\
Qwen3.8-Max & \multicolumn{3}{c}{32{,}768 / 2{,}048 \quad vs.\quad 32{,}768 / 2{,}048, trace in prompt} \\
GPT-5.6 Sol & \multicolumn{3}{c}{reasoning effort high \quad vs.\quad same, trace in prompt} \\
\bottomrule
\end{tabular}
\end{table}

Table~\ref{tab:hyperparams} lists the main training and evaluation settings. Training
uses verl~\citep{verl} with FSDP2~\citep{zhao2023fsdp} and vLLM~\citep{kwon2023vllm} rollouts. Every reinforcement-learning run,
ours and the GRPO and EA-GRPO controls, shares the upper block and differs only in how
the reward is turned into advantages. The model samples with thinking enabled, and the
thinking budget is enforced during both training and evaluation. Checkpoints are saved and
scored on the validation split every 25 steps. Evaluation decodes one sample per task with
the same minimal-edit prompt used in training (Appendix~\ref{app:prompts}). GT Debug is
given a larger budget so that the model has room to decide that no edit is needed.

\subsection{Baselines}
\label{app:baselines}

\paragraph{Rejection-sampling SFT and DPO.}
Both baselines are self-distilled from the base model they train, with no stronger teacher. The
base model samples 16 repairs per training task (temperature 1.0, top-$p$ 1.0, a 4{,}096-token
thinking budget and 6{,}144 tokens in total), and samples that fail the unit tests are dropped.
For SFT the passing samples are ranked, by edit precision at 4B and by edit $F_1$ at 9B, and the
top two per task are kept, giving 7{,}116 targets at 4B and 6{,}821 at 9B; the model is fine-tuned
for two epochs at learning rate $5\times10^{-6}$. For DPO the same 16 samples are ranked by
$\mathrm{unit}+\tfrac12 F_1$ and paired as first against fifteenth and second against sixteenth,
giving 7{,}501 pairs at 4B and 7{,}062 at 9B; the sigmoid DPO loss is trained for one epoch at
$\beta=0.1$ and learning rate $5\times10^{-7}$, starting from the base model, which also serves as the
reference policy.

\paragraph{EA-GRPO.}
EA-GRPO~\citep{qimengprepair} keeps the unit-test reward and subtracts an edit-size penalty from it inside
rollout groups that already pass often enough. For a group of $G$ rollouts of one buggy program, let $u_i\in\{0,1\}$ be
the unit-test outcome and $D_i$ the number of lines changed between the input and rollout $i$. The penalty is switched
on by a group-accuracy gate $T=\mathbf{1}[\tfrac1G\sum_i u_i\ge\alpha]$ and, among the passing rollouts, is the
sigmoid of the standardized edit size, $P_i=\sigma\big((D_i-\mu_c)/\sigma_c\big)$ with $\mu_c,\sigma_c$ the mean and
standard deviation of $D$ over the passing subset. The reward is $R_i=u_i\,(1-T\beta P_i)$: a failing rollout scores
0, a passing rollout scores between $1-\beta$ and 1 according to how much it edits relative to the other passing
rollouts, and the group is then normalized as in standard GRPO. We use $\alpha=0.8$ and $\beta=0.05$, the setting their
ablation on Qwen2.5-Coder-7B selects: they report that a low $\alpha$ penalizes every group and costs correctness, a high
$\alpha$ never activates the penalty, and a larger $\beta$ lowers pass@1. Their edit distance normalizes by the
program length; since the length is shared within a group, it cancels under the standardization and we use the
changed-line count directly. Data, prompt, group size and optimizer settings are those of Section~\ref{sec:exp-setup}.

\paragraph{RECAP.}
RECAP~\citep{recap} repairs in two stages: a generator proposes a fix, and a separately trained refiner
rewrites it into a correct, minimal patch. We use each base model as its own generator and fine-tune the
4B and 9B models as refiners. Refiner inputs are the problem, the buggy code and a candidate fix sampled
from the 4B or 9B base model on a \pdb{} training task; targets are a correct minimal fix, either the
ground-truth solution or the most precise passing sample. The training set mixes three cases in a
$3{:}1{:}1$ ratio: a correct but over-edited candidate that must be trimmed (correct $\to$ correct), an
incorrect candidate that must be repaired (incorrect $\to$ correct), and a candidate that is already a
correct minimal fix and must be returned as is. Each target is paired with a reasoning trace generated by
DeepSeek V4.1 Flash~\citep{deepseekai2026deepseekv41flash}, and the 3{,}761 examples that pass every filter
(3{,}381 train, 380 validation) are used to fine-tune for six epochs. We also trained the refiner with DPO,
on on-policy pairs and on pairs built from DeepSeek outputs, but neither improved on SFT, so we report the
SFT refiner. At evaluation the refiner takes the same-size base model's one-shot output on each task as its
candidate and answers under the refine prompt of Appendix~\ref{app:prompt-refine}.

\FloatBarrier
\section{Prompts}
\label{app:prompts}

Every prompt is a chat exchange with a fixed system message and a per-task user message; braces mark the fields
filled in for each task. The unit tests are never shown to the model.

\subsection{Minimal-edit prompt}
\label{app:prompt-minimal}
Used for RL training, for the SFT and DPO sampling, and for every single-turn evaluation unless stated otherwise
(\pdb{} Test, GT Debug in Table~\ref{tab:main}, \pdb{}-Wild and the frontier runs). The GT Debug probe uses the same
prompt with the bug-free program in place of the buggy code, so the prompt still asserts that the code contains
errors.

\noindent System message:
\begin{promptbox}
Debug the given Python code. ONLY fix the bugs. Make minimal edits.
Do NOT generate a new solution based on the problem description.
Do NOT reformat lines that are already correct.
Do NOT edit or add any comments.

Your response should include:
- A self-contained, corrected Python implementation, with minimal edits.
\end{promptbox}

\noindent User message:
\begin{promptbox}
Problem description:
{task_prompt}

Buggy code:
```python
{buggy_code}
```
\end{promptbox}

\subsection{Refine prompt (RECAP)}
\label{app:prompt-refine}
Used to train and evaluate the RECAP refiner (Appendix~\ref{app:baselines}); the candidate fix is the base
model's output under the minimal-edit prompt.

\noindent System message:
\begin{promptbox}
You are given buggy Python code and a candidate fix for it. Produce the correct fix with minimal edits to the buggy code.
Keep the candidate's changes that are needed to fix the bugs, drop every change that is not needed, and correct the candidate if it is wrong or incomplete. If the candidate is already a correct minimal fix, return it unchanged.
Do NOT generate a new solution based on the problem description.
Do NOT reformat lines that are already correct.
Do NOT edit or add any comments.

Your response should include:
- A self-contained, corrected Python implementation, with minimal edits to the buggy code.
\end{promptbox}

\noindent User message:
\begin{promptbox}
Problem description:
{task_prompt}

Buggy code:
```python
{buggy_code}
```

Candidate fix:
```python
{candidate_solution}
```
\end{promptbox}

\subsection{Abstain-or-fix prompt}
\label{app:prompt-abstain}
Used for the GT Debug restraint experiment (\Tabref{tab:gt-prompt}). It keeps the three ``Do NOT'' constraints of
the minimal-edit prompt but makes the bug investigation an explicit first step and licenses an unchanged answer; the
user message labels the program ``Code:'' rather than ``Buggy code:''.

\noindent System message:
\begin{promptbox}
First, investigate whether the given Python code contains bugs. If it contains no bugs, do NOT make any changes and return the code exactly as given. Otherwise, ONLY fix the bugs. Make minimal edits.
Do NOT generate a new solution based on the problem description.
Do NOT reformat lines that are already correct.
Do NOT edit or add any comments.

Your response should include:
- A self-contained Python implementation: the original code unchanged if it contains no bugs, otherwise the corrected code with minimal edits.
\end{promptbox}

\noindent User message:
\begin{promptbox}
Problem description:
{task_prompt}

Code:
```python
{code}
```
\end{promptbox}

\subsection{Chain-of-thought transfer}
\label{app:prompt-transfer}
The receiver is given exactly the chat prompt the donor answered, that is the minimal-edit system and user messages
above, and the donor's think block is spliced into the receiver's assistant turn over a raw completions endpoint.
In the ``answers at once'' setting the block is closed immediately and the receiver writes the repair:
\begin{promptbox}
<think>
{donor think block}
</think>
\end{promptbox}

\noindent In the ``continues thinking'' setting the block stays open after the donor trace; the receiver adds its own
reasoning up to a thinking budget, after which the block is closed and the receiver answers:
\begin{promptbox}
<think>
{donor think block}
{receiver continues thinking, up to its budget}
</think>
\end{promptbox}

\noindent Where the API cannot prefill an assistant turn (Qwen3.8-Max, GPT-5.6 Sol), the donor trace is appended to the
user message instead and the receiver reasons on its own:
\begin{promptbox}
Problem description:
{task_prompt}

Buggy code:
```python
{buggy_code}
```

The thinking starts with:
{donor think block}
\end{promptbox}

\subsection{Agent tool prompt}
\label{app:prompt-agent}
The agentic evaluation (\figref{fig:agentic}) runs each task as a five-turn tool loop over an in-memory copy of the
buggy file, with the tools \texttt{view}, \texttt{str\_replace\_editor} and \texttt{submit} serialized in the
SWE-agent plain-text format; the \texttt{run\_tests} tool exists in the harness but is disabled in the reported runs.
Exactly one tool call is applied per turn. \texttt{view} returns the requested lines with line numbers;
a successful \texttt{str\_replace\_editor} call returns ``Edit applied.'' with the surrounding lines, and a
non-unique or non-matching \texttt{old\_str} returns ``Edit failed:'' with the reason and leaves the file unchanged.
The session ends at \texttt{submit} or when the turn budget runs out, and the current file is scored.

\noindent System message:
\begin{promptbox}
You are debugging a single Python file. ONLY fix the bugs. Make minimal edits.
Do NOT rewrite the file from scratch. Do NOT reformat lines that are already correct. Do NOT add or edit comments.

You work by emitting tool calls. Emit EXACTLY ONE tool call per reply, using this plain-text format. Anything outside the function block is ignored. If a reply contains no tool call, nothing is applied and you will be asked to try again. The session ends only when you call submit (or the turn budget runs out).
Do NOT answer with the corrected file as a ```python code block or any other prose @\textemdash@ that is NOT a tool call and is ignored. Every change must go through a str_replace_editor tool call.

Available tools:

view @\textemdash@ show the current file with line numbers. view_range [start, end] (1-based, inclusive) is optional; include it to show only that region, or omit it to show the whole file:
    <function=view>
    <parameter=view_range>[10, 30]</parameter>
    </function>

str_replace_editor @\textemdash@ replace a unique snippet. old_str must match the CURRENT file EXACTLY, including whitespace and indentation, and must occur EXACTLY ONCE; otherwise the file is left unchanged and you get an error. Do NOT include the line-number prefixes shown by view in old_str. Edits accumulate across turns.

    <function=str_replace_editor>
    <parameter=old_str>    for i in range(n):
        total += i</parameter>
    <parameter=new_str>    for i in range(1, n):
        total += i</parameter>
    </function>

To INSERT new lines (nothing is being removed), pick a unique anchor line next to the insertion point, then repeat the anchor in new_str:

submit @\textemdash@ finish; your current file is scored.
    <function=submit>
    </function>

(The blocks above are format examples, not the actual edit to make.)
\end{promptbox}

\noindent First user message (the file is shown with line numbers):
\begin{promptbox}
Problem description:
{task_prompt}

Current contents of the file (line numbers shown for reference only @\textemdash@ do not include them in old_str):
```python
     1	{buggy_code line 1}
     2	{buggy_code line 2}
     3	...
```

Fix the bug(s) with the fewest possible str_replace edits, then call submit.
\end{promptbox}

\noindent A reply with no tool call is answered with this user message and the turn is retried:
\begin{promptbox}
Your last reply contained no tool call, so nothing was applied. A ```python code block or prose answer is ignored @\textemdash@ you can only change the file through a tool call in the exact format, e.g.
<function=str_replace_editor>
<parameter=old_str>...exact current line(s)...</parameter>
<parameter=new_str>...replacement...</parameter>
</function>
Translate your intended fix into one such tool call now, or call <function=submit></function> if the file is already correct.
\end{promptbox}

\FloatBarrier
\section{Limitations}
\label{sec:limitations}

\paragraph{All training and evaluation are in Python.} The benchmarks we build on are Python: we train on
\pdb{} tasks derived from BigCodeBench and LiveCodeBench and evaluate on these, on SWE-smith repositories
(\pdb{}-Wild), SWE-bench Verified, HumanEval+, CruxEval and the Python subset of DebugBench. Nothing in the
reward is specific to Python: precision and recall are computed on line-level diffs against a reference fix,
and semantic verification only executes the tests. Extending \pc{} to another language therefore requires
buggy programs with reference fixes and a test harness in that language, such as the C++ and Java splits of
DebugBench~\citep{debugbench}, rather than a change to the method; we have not run this experiment.

\paragraph{The reward needs a reference repair.} Precision and recall are measured against a
ground-truth fix (\Secref{sec:method:roles}), so every training task needs one. The reference can come
from the benchmark itself, as in \pdb{}; from a human-written fix, such as the developer patches of
SWE-bench Verified~\citep{swebenchverified}; or, where neither exists, from a model-generated fix that
passes the unit tests.

\paragraph{Scoring the edit costs training time.} Computing precision and
recall, including the semantic verification of each edit block, adds about
5--7\% to the wall-clock time of unit-test-only training. Caching semantic
verification results reduces this cost, and it is amortized further as
training scales to larger groups, longer rollouts or larger models, where
generation and the policy update dominate.

\end{document}